\documentclass[a4paper,11pt]{article}
\pdfoutput=1 

\usepackage{jheppub} 

\usepackage[T1]{fontenc} 

\usepackage{mathrsfs}
\usepackage{amsmath}
\usepackage{dsfont}
\usepackage{hyperref}
\usepackage{enumitem}

\usepackage{slashed} 

\usepackage{graphicx}
\usepackage{dcolumn}
\usepackage{bm}
\usepackage{upgreek}
\def\p{\textbf{p}}

\def\0{\textbf{0}}
\def\P{\textbf{P}}
\def\J{\textbf{J}}
\def\K{\textbf{K}}
\def\x{\textbf{x}}

\renewcommand\thefigure{\arabic{figure}}

\newcommand{\dual}[1]{\overset{\:{}^{^{{{\neg}}}}}{\smash[t]{#1}}} 

\newcommand\varpm{\mathbin{\vcenter{\hbox{%
  \oalign{\hfil$\scriptstyle+$\hfil\cr
          \noalign{\kern-.3ex}
          $\scriptscriptstyle({-})$\cr}%
}}}}
\newcommand\varmp{\mathbin{\vcenter{\hbox{%
  \oalign{$\scriptstyle({+})$\cr
          \noalign{\kern-.3ex}
          \hfil$\scriptscriptstyle-$\hfil\cr}%
}}}}

\title{\boldmath
Wigner multiplets in QFT: dark sector and $CPT$-violating scenarios}
\author[a]{Cheng-Yang Lee}
\affiliation[a]{Center for theoretical physics, College of Physics, Sichuan University, \\Chengdu, 610064, China } 
\emailAdd{cylee@scu.edu.cn}
\author[b]{Ruifeng Leng}
\affiliation[b]{Department of Physics and Center for Field Theory and Particle Physics, \\Fudan University, Shanghai 200438, China}
\emailAdd{lruifeng@fudan.edu.cn}
\author[c]{Siyi Zhou}
\emailAdd{siyi@cqu.edu.cn} 
\affiliation[c]{Department of Physics and Chongqing Key Laboratory for Strongly Coupled Physics,
Chongqing University, Chongqing 401331, China} 
\abstract{
The classification of elementary particles based on unitary irreducible representations of the Poincar\'{e} group has been a cornerstone of modern Quantum Field Theory (QFT). 
While the Standard Model (SM) does not inherently include Dark Matter (DM), any fundamental DM candidate should still conform to this classification or its extensions. 
Eugene P. Wigner introduced a class of nontrivial representations characterized by an additional discrete degree of freedom, known as the Wigner degeneracy. 
In this work, we systematically investigate the QFT of such Wigner multiplets, particularly focusing on the massive spin-1/2 fermion. 
We construct a theoretical framework where the two-fold Wigner spinor fields, $\psi_{\pm\frac{1}{2}}(x)$, form a doublet representation. We analyze their transformation properties under discrete symmetries (e.g., charge-conjugation $C$, spatial parity $P$, and time-reversal $T$), revealing novel mixing effects due to the Wigner degeneracy and an emergent accidental $U(2)$ global symmetry. 
Furthermore, we explore the Yukawa interactions involving the Wigner doublets, showing that such interactions generally violate the $CPT$ invariance. 
We also study gauge theories within the Wigner framework, where the physical Wigner doublet naturally leads to exotic phenomenological consequences beyond the SM, including phase transitions.
These results provide new insights into the possible role of the Wigner-degenerate states in fundamental physics, particularly in the dark sector.  
}

\begin{document} 
\maketitle
\flushbottom

\section{Introduction}
\label{sec:intro}

The paradigm that elementary particles are described by unitary irreducible representations of the Poincar\'{e} group has been remarkably successful~\cite{1939AnMat..40..149W}. As long as the Poincar\'{e} symmetry holds and quantum gravitational effects remain negligible, this paradigm is expected to remain valid. Although the SM does not originally include DM, there is no reason to assume that DM should not be described by the representations of the Poincar\'{e} group and its extension. The key question is: what representation does DM furnish?

In the search for physics beyond the SM, to ensure that no stones are left unturned, it is essential to explore all unitary irreducible representations of the extended Poincar\'{e} group and their corresponding QFTs. 
One such representation of particular interest, which may serve as a potential DM candidate, was discovered by Eugene P. Wigner~\cite{wigner1964unitary}. When the continuous Poincar\'{e} group is extended to include discrete transformations, Wigner demonstrated the existence of a class of nontrivial representations in which a one-particle state acquires an additional degree of freedom beyond the conventional attributes such as four-momentum and spin projection. 
While Wigner did not further develop this idea, Steven Weinberg later examined these representations and referred to the corresponding states as \textit{degenerate multiplets}~\cite[App.~2C]{Weinberg:1995mt}. In this work, we denote such an on-shell one-particle state by $|\p,\sigma,n\rangle$ where $\p$ denotes the three-momentum, $\sigma$ is the spin projection, and $n$ represents the additional discrete degree of freedom, characterizing the \textit{Wigner degeneracy}. The state is degenerate in the sense that $n$ remains invariant under continuous Lorentz transformations and spacetime translations. In contrast, it is fundamentally affected by discrete transformations, which map $|\p,\sigma,n\rangle$ into a superposition of Wigner-degenerate states with properly transformed $\p$ and $\sigma$.

Weinberg provided a brief introduction to the discrete transformations in the Wigner framework. However, he did not proceed to develop the corresponding QFT, due to the practical consideration that \textit{no examples are known of particles that furnish unconventional representations of inversions}. 
While this statement holds within the SM, it is crucial to realize that approximately 25\% of the total energy-matter content of the observable universe exists in the form of DM. Thus, one should remain open to the possibility that such unconventional representations may describe particles yet to be discovered. 
This nontrivial representation could play a fundamental role in the dark sector, governing both the self-interactions of DM particles and their couplings to the SM matter. From a theoretical perspective, our understanding of the continuous and discrete symmetries of Wigner multiplets and the associated quantum fields remains incomplete.

The primary objective of this work is to explore the Wigner multiplets and their associated QFT within a straightforward and intuitive construction. We focus on the massive spin-1/2 Wigner-degenerate fermions in order to construct DM fields, that may serve as an alternative to Majorana fermions~\cite{Weinberg:1964cn,Weinberg:1995mt}. The Wigner degeneracy parameter $n$ can take discrete values $-w,\cdots,w$ where $w$ is an arbitrary positive integer or half-integer. 
For simplicity, we consider the two-fold Wigner case $w=\frac{1}{2}$, $n=\pm\frac{1}{2}$, introducing the Wigner doublet. A one-particle state in this framework possesses four degrees of freedom: two from spin projections $\sigma=\pm\frac{1}{2}$ and two corresponding to $n=\pm\frac{1}{2}$. 
The corresponding quantum fields must incorporate both degrees of freedom while satisfying the causality condition and Poincar\'{e} invariance. 
Since the Wigner degeneracy is independent of spacetime transformations, we construct a pair of quantum fields $\psi_{n}(x)$, $n=\pm\frac{1}{2}$ corresponding to the Wigner doublet. 
These fields are required to be causal and Lorentz covariant. While Poincar\'{e} transformations preserve the Wigner degeneracy, discrete inversions, including the charge-conjugation, can introduce nontrivial mixing between the Wigner-degenerate states. To fully capture the physical nature of the two-fold Wigner degeneracy, it is crucial to effectively combine the two fields. This raises the key question: \textit{how should they be combined?} We consider two reasonable approaches:
\begin{enumerate}
    \item Doublet construction: $\Psi(x)\equiv\left[\begin{matrix}
        \psi_{+\frac{1}{2}}(x) \\
        \psi_{-\frac{1}{2}}(x)
    \end{matrix}\right]$;
    \item Superposition over the Wigner degeneracy: $\lambda(x)\equiv \dfrac{1}{\sqrt{2}} \left[ \psi_{+\frac{1}{2}}(x) + \psi_{-\frac{1}{2}}(x) \right]$.
\end{enumerate}
In this work, we primarily focus on the first approach --- the doublet construction.
The central task is then to explore the action of discrete transformations $C$, $P$, and $T$ on both Wigner-degenerate (anti-)particle states and quantum fields, along with their bilinear forms.
These actions generally differ from those in the conventional QFT (CQFT) due to the mixing of Wigner degeneracy, leading to modified symmetry properties.

We are simultaneously exploring the second approach --- the superposition framework --- as a separate research project. 
In the literature, this framework has been imposed in the theory of mass-dimension-one fields, which presents a shift in the standard paradigm~\cite{Ahluwalia:2020jkw,Ahluwalia:2022zrm,ahluwalia_2019,Ahluwalia:2022ttu,Ahluwalia:2022yvk}.
These fields have been applied across various phenomenological domains, including cosmology~\cite{Basak:2014qea,Boehmer:2006qq,Boehmer:2007dh,Boehmer:2008ah,Boehmer:2008rz,Boehmer:2009aw,Boehmer:2010ma,HoffdaSilva:2014tth,Pereira:2014wta,S:2014woy,Pereira:2014pqa,Pereira:2016emd,Pereira:2016eez,Pereira:2017efk,Pereira:2017bvq,BuenoRogerio:2017zxf,Pereira:2018xyl,Pereira:2018hir,Pereira:2020ogo,Pereira:2021dkn,Lima:2022vrc}, braneworld models~\cite{Jardim:2014xla,Dantas:2015mfi,Zhou:2017bbj,ZhouZhouXiangNan:2018het,Sorkhi:2018jhy,MoazzenSorkhi:2020fqp} and models of self-interacting DM~\cite{Dias:2010aa,Agarwal:2014oaa,Alves:2014kta,Alves:2014qua,Alves:2017joy,Moura:2021rmf}.
While the theoretical foundations of the superposition field (i.e., the Elko field) have been extensively studied, most works have primarily focused on ensuring their compatibility with causality and rotational symmetries~\cite{Ahluwalia:2004sz,Ahluwalia:2004ab,Ahluwalia:2022yvk,Ahluwalia:2023slc}.
However, a systematic methodology for their construction remains unclear. In the next stage of our research, we aim to fill this gap. This effort is expected to provide deeper insights into the underlying structure of Wigner degeneracy and its potential implications in physics beyond the SM.

This paper is organized as follows. 
In Section~\ref{sec:wd}, we begin with a quick review of the standard representations in CQFT, followed by an introduction to the concept of Wigner degeneracy. We then present the general representations of the continuous Lorentz group and discrete inversions acting on Wigner multiplets.
In Section~\ref{sec:Spin_m_1}, we conduct a detailed study of Wigner doublets, which correspond to the quantum fields associated with the two-fold Wigner-degenerate spinors. We explicitly derive the transformation properties of both one-particle states and their corresponding quantum fields under charge-conjugation $C$, parity $P$, and time-reversal $T$. 
In addition, we discuss the fundamental conditions required for constructing a nontrivial QFT of physical Wigner doublets.
Section~\ref{sec:cpt_thm} focuses on the product $CPT$ transformation. While the $CPT$ invariance is always valid in CQFT, the presence of Wigner degeneracy introduces additional structures, leading to new and nontrivial physical phenomena.
In Section~\ref{sec:can_W_1}, we develop the canonical Lagrangian formalism for free Wigner doublets and analyze their global symmetries. The discussion then extends to interacting theories in Section~\ref{sec:W_inter}, where we explore possible interactions of Wigner-degenerate fields, particularly in the context of theoretical dark matter candidates.
The results presented in this work suggest novel phenomenological implications beyond the SM, motivating further investigation into the role of Wigner degeneracy in fundamental physics and cosmology.



\section{The Wigner degeneracy}
\label{sec:wd}

In Minkowski spacetime, elementary particles are described by the irreducible unitary representations of the Poincar\'{e} group~\cite{1939AnMat..40..149W}. Since we are dealing with physical states, the relevant representations correspond to massive and massless particles, both of which possess positive-definite mass and energy.  
That is, for a physical particle with four-momentum $p^{\mu} = (E_{\p}, \p)$, we impose the conventions $p^{\mu}p_{\mu} = m^2 \geq0$ and $p^{0} = E_{\p} \geq0$.
The representations of Poincar\'{e} group can be classified by the eigenvalues of its two Casimir invariants: $P_{\mu} P^{\mu}$ and $W_{\mu} W^{\mu}$, where 
$P^{\mu}$ is the four-momentum operator and
$W^{\mu}$ is the Pauli-Lubanski vector. 
The two corresponding eigenvalues are $m^{2}$ and $m^{2}j(j+1)$, where $m$ represents the mass of the particle, and $j=0,\frac{1}{2},1,\cdots$ is its spin. For massless particles, the spin is replaced by helicity $\pm j$. 
An intriguing class of nontrivial representations emerges when the
Poincar\'{e} group is extended to include discrete symmetries. 
Originally discovered by Wigner~\cite{wigner1964unitary} and later developed by Weinberg~\cite{Weinberg:1995mt},
these representations introduce an additional degeneracy $n$ in the Hilbert space, referred to as the \textit{Wigner degeneracy}.
In Appendix~\ref{sec:common_rep_1}, we provide a concise review of the standard representations for massive particles, establishing the key formulations necessary to set the stage for our discussion.

In 1964, Wigner introduced a novel class of nontrivial representations for inversions $P$ and $T$, 
in which quantum states acquire additional internal degrees of freedom and transform in a more intricate pattern~\cite{wigner1964unitary}. 
In this framework, the actions of $P$ and $T$ can map a given state to a superposition of degenerate states. Hence, the representations of spacetime inversions are not necessarily faithful.
Weinberg clarified this representation particularly on one-particles and referred to them as \textit{degenerate multiplets}~\cite[App.~2C]{Weinberg:1995mt} under particular physical requirements instead of imposing some of Wigner's limiting assumptions~\footnote{Wigner assumed the square of inversion operators are proportional to the unit operator; in other words, the (projective) representations of inversions on spacetime are faithful.}.
Following Weinberg’s prescriptions, we introduce one-particle states characterized not only by momentum $\p$ and spin-projection $\sigma$, but also by an additional discrete quantum number referred to as the Wigner degeneracy, labeled by $n$. 
This additional index expands the notation of one-particle state as $|\p,\sigma,n\rangle$, 
where $\p$ is the three-momentum, $\sigma = -j, \cdots, +j$ denotes the spin-projection, and $n= -w,\cdots,+w$ labels the Wigner degeneracy. 
These states form an orthonormal basis in the Hilbert space with the Lorentz invariant normalization:
\begin{equation}
    \langle \p^{\prime},\sigma^{\prime}, n' |\p,\sigma, n \rangle 
    = 2 E_{\p} \left(2\pi \right)^{3} \delta^{(3)} \left( \p^{\prime} - \p \right) \delta_{\sigma^{\prime} \sigma} \delta_{n' n} \, .
\label{eq:norm_1p_gen_2}
\end{equation}
The associated creation operator is denoted by $ a_{n}^{\dag}(\p, \sigma)$, satisfying the canonical (anti-)commutation relations:
\begin{equation}
\bigl[a_{n}(\p,\sigma),a_{n'}^{\dag}(\p^{\prime},\sigma')\bigr]_{\mp} 
= (2\pi)^{3} \delta^{(3)} \left( \p^{\prime} - \p \right) \delta_{\sigma^{\prime} \sigma} \delta_{n' n} \, ,
\label{can_comm_2}
\end{equation}
and generating the one-particle state from the vacuum via
\begin{equation}
    |\p,\sigma, n \rangle \equiv \sqrt{2 E_{\p}} \,  a_{n}^{\dag}(\p, \sigma) |0\rangle \, .
\label{eq:deg_op}
\end{equation}
The vacuum $|0\rangle$ is defined such that it can be annihilated by all the annihilation operators $a_{n}(\boldsymbol{p}, \sigma)$
\begin{align}
 a_{n}(\boldsymbol{p}, \sigma) |0\rangle  = 0 \, .
\label{eq:vac_W_a_1}
\end{align}
For the remainder of this section, we will review how these Wigner-degenerate one-particle states transform under Lorentz transformations and discrete inversions. The extension of these results to quantum fields will be presented in the following section.

Lorentz transformations are assumed to act trivially on the Wigner degeneracy label.  
That is, under a Lorentz transformation $\Lambda$, the massive one-particle states transform in the same way as in the standard representations, with the Wigner degeneracy index $n$ remaining unchanged:
\begin{equation}
U(\Lambda)|\p,\sigma, n \rangle
=\sum_{\sigma'}D^{(j)}_{\sigma'\sigma}(W(\Lambda,p))|\p_{\Lambda},\sigma', n \rangle \,.\label{eq:Lorentz_trans}
\end{equation}
This explicitly shows that each fixed value of $n$ defines an invariant subspace of the Lorentz group.
Consequently, the Hilbert space of massive one-particle states with mass $m$ can be decomposed into $2w+1$ Lorentz invariant subspaces via the direct sum 
\begin{align}
\mathscr{H} = \displaystyle{\bigoplus^{+w}_{n=-w} V_{n} } \, , 
\quad
V_{n} \equiv \left\{ |\p,\sigma,n\rangle \ | \ \p \in \mathbb{R}^3, \ \text{and} \ \sigma=-j,\cdots, +j \right\}
\, .
\end{align}
However, what can potentially mix the Wigner-degenerate multiplets is the action of discrete inversions.
In particular, the parity $P$ and the time-reversal $T$ act on the one-particle states as 
\begin{align}
    P|\p,\sigma,n\rangle&=\sum_{n'}D_{n'n}(\mathscr{P})|-\p,\sigma,n'\rangle \, ,
    \label{eq:P_state_W_1}\\
    T|\p,\sigma,n\rangle&=(-1)^{j-\sigma}\sum_{n'}D_{n'n}(\mathscr{T})|-\p,-\sigma,n'\rangle \, ,
    \label{eq:T_state_W_1}
\end{align}
where $D_{n'n}(\mathscr{P})$ and $D_{n'n}(\mathscr{T})$ are unknown matrices except that they are unitary. 
Hence, the normalization of physical states~\eqref{eq:norm_1p_gen_2} is preserved. 
The inversion operators $P$ and $T$ are required to obey the same algebra with the Poincar\'{e} generators as in CQFT. 
It is important to note that Wigner originally imposed additional, physically unmotivated constraints by assuming that the squared actions of inversion operators are proportional to the identity operator, i.e., 
$P^{2} = \zeta_{P} \boldsymbol{1}$~\footnote{To be clear, here `$\boldsymbol{1}$' is the identity operator on the Hilbert space.} and $T^{2} = \zeta_{T} \boldsymbol{1}$ with constants $\zeta_{P,T}$.
If we impose Wigner's assumption, there will be an extra constraint $\zeta_{T} = \pm 1$ that can be proved by the antiunitarity and antilinearity of $T$~\footnote{Since $T$ is antiunitary and antilinear, $T^{2}$ is unitary so that $\zeta_{T}$ is a phase at most. Then, the antilinearity of $T^{(\dag)}$ implies $\zeta_{T} = \zeta_{T}^{*}$ (real), demonstrated by
\begin{align}
\zeta_{T} T^{\dag}
= T^{2} T^{\dag}
= T^{\dag} T^{2}
= \zeta_{T}^{*} T^{\dag}
\, ,
\nonumber
\end{align}
where we have used the property of antiunitary $T$ that 
$T^{\dag} T = T T^{\dag} = \boldsymbol{1}$. Thus, we can conclude that $\zeta_{T} = \pm 1$ and $\zeta_{T}$ are highly constrained under Wigner's assumption. In contrast, we can only argue $\zeta_{P}$ as a phase factor because $P$ is unitary and linear.
}.

If there exists a basis that can diagonalize the two inversion matrices $D_{n'n}(\mathscr{P})$~\eqref{eq:P_state_W_1} and $D_{n'n}(\mathscr{T})$~\eqref{eq:T_state_W_1} simultaneously, then the Wigner multiplet reduces to a trivial replication of the standard fermionic representation, with each degenerate copy transforming identically under both parity and time-reversal.
However, in general, such simultaneous diagonalization is not possible. 
While the unitary and linear matrix $D_{n'n}(\mathscr{P})$ can always be diagonalized, the time-reversal matrix $D_{n'n}(\mathscr{T})$ is more subtle due to the antiunitary and antilinear nature of $T$.
In most cases, we can only block-diagonalize $D_{n'n}(\mathscr{T})$ with the block either a phase factor or a $2 \times 2$ matrix of phases
\begin{align}
\left[\begin{matrix}
        0 & e^{i\frac{\phi}{2}} \\
        e^{-i\frac{\phi}{2}} & 0
\end{matrix}\right]
\, , \ \ \text{with} \ \ 
\phi \in \mathbb{R}
\, .
\label{UT_W_B_1}
\end{align}
Even if we successfully diagonalize $D_{n'n}(\mathscr{T})$ by accident under very limited situations, 
the associated basis transformation does not generally diagonalize $D_{n'n}(\mathscr{P})$ simultaneously (see \cite[App.~2C]{Weinberg:1995mt} for more details). 
The actions of two inversion operators on the annihilation and creation operators follow directly from Eqs.~\eqref{eq:P_state_W_1} and \eqref{eq:T_state_W_1}:
\begin{align}
P a_{n}^{\dag}(\p, \sigma) P^{-1} &= 
\sum_{n'}D_{n'n}(\mathscr{P}) a_{n'}^{\dag}(-\p, \sigma) \, ,
\label{eq:P_part_W_1} \\
P a_{n}(\p, \sigma) P^{-1} &= 
\sum_{n'}D_{n'n}^{*}(\mathscr{P}) a_{n'}(-\p, \sigma) \, ,
\label{eq:P_part-_W_1}
\\
T a_{n}^{\dag}(\p, \sigma) T^{-1} &= 
(-1)^{j-\sigma}\sum_{n'}D_{n'n}(\mathscr{T}) a_{n'}^{\dag}(-\p, -\sigma) \, ,
\label{eq:T_part_W_1} \\
T a_{n}(\p, \sigma) T^{-1} &= 
(-1)^{j-\sigma}\sum_{n'}D_{n'n}^{*}(\mathscr{T}) a_{n'}(-\p, -\sigma) \, .
\label{eq:T_part-_W_1}
\end{align}
In the following sections, we will study QFT in the presence of Wigner degeneracy. For free fields, we will construct the most general formulation that preserves both the locality condition and the Lorentz covariance. 
In particular, we will investigate whether these fields exhibit distinct kinematic features compared to their standard counterparts. On the phenomenological aspect, we will explore possible interaction structures that could reveal observable signatures of the underlying Wigner degeneracy.

\section{The two-fold spinor fields}
\label{sec:Spin_m_1}



In this section, we establish a fundamental framework for quantum spinor fields describing a massive spin-1/2 fermion doublet, which has internal degrees of freedom, including the usual spin projection $\sigma=\pm\frac{1}{2}$ and, in particular, the two-fold Wigner degeneracy $w=\frac{1}{2}, \ n=\pm\frac{1}{2}$. 
Thus, a Wigner-degenerate fermion must be an excitation of a quantum spinor field that incorporates both the spin projection and the Wigner degeneracy as fundamental components of its degrees of freedom.

\subsection{Construct the Wigner-degenerate fields from the Lorentz symmetry}

Following the methodology in CQFT, we construct two Dirac spinor fields in the doublet form for the massive two-fold Wigner-degenerate fermions~\footnote{One may be naively tempted to construct a quantum field via the linear combination
$\psi(x) \equiv \frac{1}{\sqrt{2}} \bigl[ \psi_{+\frac{1}{2}}(x) + \psi_{-\frac{1}{2}}(x) \bigr]$ 
instead of the doublet~\eqref{eq:doublet} and proceed to quantize it with the Lagrangian density $\mathcal{L}'=\bar{\psi}(i\gamma^{\mu}\partial_{\mu}-m)\psi$. However, using the explicit configurations of spinor fields in this section, canonical calculation reveals that $\mathcal{L}'$ does not yield the correct free Hamiltonian (see Section~\ref{sec:L_W2_free}). To quantize $\psi(x)$, a different strategy similar to those proposed in Ref.~\cite{Ahluwalia:2022yvk,Ahluwalia:2023slc} is necessary.}
\begin{equation}
    \Psi(x)\equiv\left[\begin{matrix}
        \psi_{+\frac{1}{2}}(x) \\
        \psi_{-\frac{1}{2}}(x)
    \end{matrix}\right] \, , 
    \label{eq:doublet}
\end{equation}
along with its Dirac dual defined as~\footnote{In matrix form $\overline{\Psi}(x)
\equiv
\bigl[\begin{matrix}
    \bar{\psi}_{+\frac{1}{2}}(x) & \bar{\psi}_{-\frac{1}{2}}(x)
\end{matrix} \bigr]$.}
\begin{equation}
    \overline{\Psi}(x)\equiv\Psi^{\dag}(x)\gamma^{0}\, .
\label{eq:doublet_D_conj}
\end{equation}
Each component $\psi_{n}(x)$, $n=\pm\frac{1}{2}$ is a Dirac causal field given by:
\begin{align}
    \psi_{n, \ell}(x)
    &=\int\frac{d^{3}p}{(2\pi)^{3}}\frac{1}{\sqrt{2E_{\p}}}\sum_{\sigma}\left[e^{-ip\cdot x}u_{n, \ell}(\p,\sigma)a_{n}(\p,\sigma)+e^{ip\cdot x} v_{n, \ell}(\p,\sigma)a^{c \dag}_{n}(\p,\sigma)\right] \, ,
    \label{eq:sf_W_20}
\end{align}
ensuring their covariance under the Poincar\'{e} group. Here, $a_{n}^{(\dag)}(\p,\sigma)$ and $a^{c(\dag)}_{n}(\p,\sigma)$ are the annihilation (creation) operators for the Wigner-degenerate particle and its associated anti-particle respectively, both with the identical mass $m$, so that the one-particle states are defined as~\footnote{We use the bracket to denote an alternative item. For instance, in Eq.~\eqref{eq:deg_op_ab}, the expression can be interpreted using either $a, a_{n}^{\dag}$ or $a^{c}, a_{n}^{c \dag}$.}
\begin{align}
|\p,\sigma, n; a^{(c)} \rangle \equiv \sqrt{2 E_{\p}} \,  a_{n}^{(c) \dag}(\p, \sigma) |0\rangle \, ,
\label{eq:deg_op_ab}
\end{align}
which are orthonormalized as
\begin{equation}
    \langle \p^{\prime},\sigma^{\prime}, n'; a^{(c)} |\p,\sigma, n; a^{(c)} \rangle 
    = 2 E_{\p} \left(2\pi \right)^{3} \delta^{(3)} \left( \p^{\prime} - \p \right) \delta_{\sigma^{\prime} \sigma} \delta_{n' n} \, .
\label{eq:norm_1p_W_2}
\end{equation}
The Wigner-degenerate states transform under the Lorentz group in the usual way as shown in Eqs.~\eqref{eq:Lorentz_part_a_1}-\eqref{eq:Lorentz_part_a-_1}, where the Wigner degeneracies remain unmixed.
The canonical fermionic quantization relations, deduced from the general case in Eqs.~\eqref{can_comm_2}-\eqref{eq:deg_op}, are given by
\begin{align}
\bigl\{a_{n}(\p, \sigma), a_{n'}^{\dag}(\p^{\prime},\sigma')\bigr\}
&=\bigl\{a^{c}_{n}(\p,\sigma), a^{c \dag}_{n'}(\p^{\prime},\sigma') \bigr\}
=(2 \pi)^{3} \delta^{(3)}(\p-\p') \delta_{\sigma\sigma'} \delta_{nn'}
\, ,
\label{eq:can_W_ab_1} \\
\bigl\{a_{n}(\p, \sigma), a_{n'}^{c}(\p^{\prime},\sigma')\bigr\}
&= \bigl\{a_{n}^{\dag}(\p, \sigma), a_{n'}^{c \dag}(\p^{\prime},\sigma')\bigr\}
=\bigl\{a_{n}(\p,\sigma), a^{c \dag}_{n'}(\p^{\prime},\sigma') \bigr\}
=0
\, .
\label{eq:can_W_ab_2}
\end{align}
The free vacuum $|0\rangle$ is defined to be the state in Hilbert space such that
\begin{align}
 a_{n}(\p, \sigma) |0\rangle = a_{n}^{c}(\p, \sigma) |0\rangle = 0 \, .
\label{eq:vac_W_ab_1}
\end{align}
In particular, the exponential phases $e^{\pm ip\cdot x}$ in Eq.~\eqref{eq:sf_W_20} ensures that the field is covariant under spacetime translations. The Lorentz covariance further requires that
\begin{equation}
    U(\Lambda) \psi_{n, \ell} (x)U^{-1}(\Lambda)
    =\sum_{\ell'}\mathscr{D}_{\ell \ell'}(\Lambda^{-1})\psi_{n, \ell'}(\Lambda x) \, , 
    \label{eq:cov_W_0}
\end{equation}
where $\mathscr{D}_{\ell\ell'}$ is the finite-dimensional Dirac spinor representation of the Lorentz group as in the common sense, satisfying the pseudo-unitary relation
\begin{align}
\gamma^{0} \mathscr{D}^{\dag}(\Lambda) \gamma^{0} = \mathscr{D}^{-1}(\Lambda) \, ,
\end{align}
and transforming the Dirac matrices $\gamma^{\mu}$ as vectors~\footnote{$\Lambda_{\nu}^{\ \mu} \equiv (\Lambda^{-1})_{\ \nu}^{\mu} = \eta_{\nu \rho} \eta^{\mu \sigma} \Lambda_{\ \sigma}^{\rho}$.}
\begin{align}
\mathscr{D}(\Lambda) \gamma^{\mu} \mathscr{D}^{-1}(\Lambda) = \Lambda_{\nu}^{\ \mu} \gamma^{\nu} \, .
\end{align}
Eq.~\eqref{eq:cov_W_0} provides the solutions for the polarizations $u_{n, \ell}(\p,\sigma)$ and $v_{n, \ell}(\p,\sigma)$ in the spinor field~\eqref{eq:sf_W_20}.
Causality condition is formulated by the two-point correlation function $\bigl\{ \psi_{n, \ell}, \psi^{\dag}_{n', \ell'} \bigr\}$, which must vanish at space-like intervals. Notice that nontrivial correlation functions are only provided by the commutators between two fields with the same Wigner degeneracy (i.e., $\bigl\{ \psi_{n, \ell}, \psi^{\dag}_{n, \ell'} \bigr\}$) due to the canonical quantization relations~\eqref{eq:can_W_ab_1}-\eqref{eq:can_W_ab_2}.
Although the parity transformation is different from the standard representation in CQFT~\eqref{eq:Up}, it's still required to transform fields at the point $x$ into a superposition of these fields at $\mathscr{P} x$, as described in Eq.~\eqref{eq:P_state_W_1}.  
Thus, without loss of generality, we can determine the Dirac spinors $u_{\pm \frac{1}{2}}(\p,\sigma)$ and $v_{\pm \frac{1}{2}}(\p,\sigma)$ in Eq.~\eqref{eq:sf_W_20} as
\begin{align}
u(\p,\sigma) &= u_{\pm \frac{1}{2}}(\p,\sigma) 
=\sqrt{\dfrac{E_{\p} + m}{2}}
\left[\begin{matrix}
    \left(\mathds{1}-\frac{\boldsymbol{\sigma} \cdot \p}{E_{\p} + m}  \right) \xi^{\sigma} \\
    \left(\mathds{1}+\frac{\boldsymbol{\sigma} \cdot \p}{E_{\p} + m}  \right) \xi^{\sigma} 
\end{matrix}\right]
=
\left[\begin{matrix}
    \sqrt{p \cdot \sigma} \xi^{\sigma} \\
    \sqrt{p \cdot \bar{\sigma}} \xi^{\sigma} 
\end{matrix}\right]
\, ,
\label{sf_u_W_20} \\
v(\p,\sigma) &= v_{\pm \frac{1}{2}}(\p,\sigma) 
=\sqrt{\dfrac{E_{\p} + m}{2}}
\left[\begin{matrix}
    \left(\mathds{1}-\frac{\boldsymbol{\sigma} \cdot \p}{E_{\p} + m}  \right) \chi^{\sigma} \\
    -\left(\mathds{1}+\frac{\boldsymbol{\sigma} \cdot \p}{E_{\p} + m}  \right) \chi^{\sigma} 
\end{matrix}\right]
=
\left[\begin{matrix}
    \sqrt{p \cdot \sigma} \chi^{\sigma} \\
    -\sqrt{p \cdot \bar{\sigma}} \chi^{\sigma} 
\end{matrix}\right]
\, ,
\label{sf_v_W_20}
\end{align}
where $u(\p,\sigma)$ and $v(\p,\sigma)$ are the solutions of the free Dirac spinors without the Wigner degeneracy and satisfy the normalization and orthogonality relations
\begin{align}
\bar{u}(\p,\sigma) u(\p,\sigma') &= -\bar{v}(\p,\sigma) v(\p,\sigma') = 2m \delta_{\sigma\sigma'} \, ,
\label{norm_uv_1} \\
\bar{u}(\p,\sigma) v(\p,\sigma') &=u^{\dag}(\p,\sigma) v(-\p,\sigma')
= \bar{v}(\p,\sigma) u(\p,\sigma') = v^{\dag}(\p,\sigma) u(-\p,\sigma') = 0 \, ,
\label{norm_uv_2}
\end{align}
with the two-component spinors
\begin{equation}
\xi^{+} =
\left[\begin{matrix}
1 \\
0
\end{matrix}\right] \, ,
\phantom{000}
\xi^{-} =
\left[\begin{matrix}
0 \\
1 
\end{matrix}\right] \, ,
\phantom{000}
\chi^{+} =
\left[\begin{matrix}
0 \\
1
\end{matrix}\right] \, ,
\phantom{000}
\chi^{-} =
\left[\begin{matrix}
-1 \\
0
\end{matrix}\right] \, .
\end{equation}

\subsection{Two inversions on the Wigner doublets}
\label{sec:PT_1}

The doublet field $\Psi(x)$~\eqref{eq:doublet} furnishes a reducible representation of the continuous Lorentz group. As a result, fermion bilinear forms constructed with the Wigner-degenerate fields in the form of $\bar{\chi}_{m} \Gamma \psi_{n}$, where $m, n = \pm \frac{1}{2}$~\footnote{$\chi$ and $\psi$ can refer to the same particle species with $\chi=\psi$. $\Gamma$ is an arbitrary $4 \times 4$ matrix.}, have the same properties of Lorentz covariance as those constructed with standard spinor fields in CQFT.
To make the doublet an irreducible representation of the extended Poincar\'{e} group,
the two discrete inversions must mix the two Wigner degeneracies. 
Two inversions can exhibit nontrivial actions on Wigner-degenerate states as
\begin{align}
    P|\p,\sigma,n; a^{(c)}\rangle&=\sum_{n'}D^{(c)}_{n'n}(\mathscr{P})|-\p,\sigma,n'; a^{(c)}\rangle \, ,
    \label{eq:P_state_W_ab}\\
    T|\p,\sigma,n; a^{(c)}\rangle&=(-1)^{\frac{1}{2}-\sigma}\sum_{n'}D_{n'n}^{(c)}(\mathscr{T})|-\p,-\sigma,n'; a^{(c)}\rangle \, .
    \label{eq:T_state_W_ab}
\end{align}
The corresponding transformations for the annihilation and creation operators follow the same form as Eqs.~\eqref{eq:P_part_W_1}-\eqref{eq:T_part-_W_1} but with the transformation matrices $D^{(c)}_{n'n}(\mathscr{P})$ and $D^{(c)}_{n'n}(\mathscr{T})$.
Since $P$ is linear and unitary, its associated matrix $D^{(c)}_{n'n}(\mathscr{P})$~\eqref{eq:P_state_W_ab} is a $2\times2$ unitary matrix, which can always be diagonalized in a suitable basis. 
In contrast, we can only block-diagonalize the matrix $D^{(c)}_{n'n}(\mathscr{T})$ for the time-reversal operator $T$~\eqref{eq:T_state_W_ab} in general.
To obtain a simple yet nontrivial result, we assume there exists a basis of the Wigner-degenerate particle and anti-particle states~\eqref{eq:deg_op_ab} that simultaneously diagonalizes $D^{(c)}_{n'n}(\mathscr{P})$~\eqref{eq:P_state_W_ab} and anti-diagonalize $D^{(c)}_{n'n}(\mathscr{T})$~\eqref{eq:T_state_W_ab} to Eq.~\eqref{UT_W_B_1},
\begin{alignat}{2}
D(\mathscr{P})&=\left[\begin{matrix}
        \eta_{+\frac{1}{2}} & 0 \\
        0 & \eta_{-\frac{1}{2}}
    \end{matrix}\right] \, ,
    \quad 
    &D^{c}(\mathscr{P})&=\left[\begin{matrix}
        \eta^{c}_{+\frac{1}{2}} & 0 \\
        0 & \eta^{c}_{-\frac{1}{2}}
    \end{matrix}\right] \, , 
    \label{mat:P_state_W_ab_2}
    \\
    D(\mathscr{T})&=\left[\begin{matrix}
        0 & e^{i\frac{\phi}{2}} \\
        e^{-i\frac{\phi}{2}} & 0
    \end{matrix}\right] \, ,
    \quad
    &D^{c}(\mathscr{T})&=\left[\begin{matrix}
        0 & e^{i\frac{\phi^{c}}{2}} \\
        e^{-i\frac{\phi^{c}}{2}} & 0
    \end{matrix}\right] \, , 
    \label{mat:T_state_W_ab_2}
\end{alignat}
where phases $\eta^{(c)}_{\pm \frac{1}{2}}$ are intrinsic parities for the (anti-)particles with Wigner degeneracies $n=\pm\frac{1}{2}$ respectively. $e^{i\frac{\phi}{2}}$ and $e^{i\frac{\phi^{c}}{2}}$ are the time-reversal phases for the Wigner-degenerate particles and the anti-particles. To be clear, we denote the $2 \times 2$ matrix element with the Wigner degeneracy $n=\pm\frac{1}{2}$ as 
\begin{equation}
    D=\left[\begin{matrix}
        D_{+\frac{1}{2},+\frac{1}{2}} & D_{+\frac{1}{2},-\frac{1}{2}} \\
        D_{-\frac{1}{2},+\frac{1}{2}} & D_{-\frac{1}{2},-\frac{1}{2}}
    \end{matrix}\right]\,.
\end{equation}
Applying these results to the Wigner-degenerate spinor fields $\psi_{n}(x)$, $n=\pm\frac{1}{2}$~\eqref{eq:sf_W_20}, we find that
\begin{align}
P\psi_{n}(x)P^{-1}
&=\gamma^{0} \int\frac{d^{3}p}{(2\pi)^{3}}\frac{1}{\sqrt{2E_{\p}}}\sum_{\sigma}
\biggl[\eta^{*}_{n} e^{-ip\cdot \mathscr{P} x}u(\p,\sigma)a_{n}(\p,\sigma) 
\nonumber\\
&\quad \quad \quad \quad \quad \quad \quad \quad \quad \quad \ \ 
- \eta^{c}_{n} e^{ip\cdot \mathscr{P} x} v(\p,\sigma)a^{c \dag}_{n}(\p,\sigma)\biggr] 
\, , \\
T\psi_{n}(x)T^{-1}
&=\gamma^{1} \gamma^{3} \int\frac{d^{3}p}{(2\pi)^{3}}\frac{1}{\sqrt{2E_{\p}}}\sum_{\sigma}
\biggl[e^{i n \phi}e^{-ip\cdot \mathscr{T}x} u(\p,\sigma) a_{-n}(\p,\sigma) 
\nonumber\\
&\quad \quad \quad \quad \quad \quad \quad \quad \quad \quad \quad \ 
+ e^{-i n\phi^{c}}e^{ip\cdot \mathscr{T}x} v(\p,\sigma) a^{c \dag}_{-n}(\p,\sigma) \biggr] 
\, ,
\end{align}
using the following identities derived from Eqs.~\eqref{sf_u_W_20}-\eqref{sf_v_W_20} 
\begin{alignat}{2}
\gamma^{0} u(\p,\sigma)&=u(-\p,\sigma) \, , 
\quad
&\gamma^{0} v(\p,\sigma)&=- v(-\p,\sigma) \, , \\
\gamma^{3} \gamma^{1} u(\p,\sigma)&=(-1)^{\frac{1}{2}-\sigma} u^{*}(-\p,-\sigma) \, , 
\quad
&\gamma^{3} \gamma^{1} v(\p,\sigma)&=(-1)^{\frac{1}{2}-\sigma} v^{*}(-\p,-\sigma) \, .
\end{alignat}
If we require the two discrete inversions to map the Wigner-degenerate fields at some point $x$ into a superposition of themselves at the corresponding point $x'$, then it is necessary to constrain the intrinsic doublet parities $\eta_{n},\eta^{c}_{n}$~\eqref{mat:P_state_W_ab_2} and the time-reversal phases $e^{i\frac{\phi}{2}}, e^{i\frac{\phi^{c}}{2}}$~\eqref{mat:T_state_W_ab_2} related as
\begin{equation}
\eta^{*}_{n} = -\eta^{c}_{n} \, ,
\quad
e^{i\frac{\phi}{2}} = e^{-i\frac{\phi^{c}}{2}} \, ,
\label{eq:PT_W_phi_2}
\end{equation}
which implies that the intrinsic spatial parities of the Wigner-degenerate particles and their anti-particles are related in the same way as in the standard representation of CQFT.
Under this condition, two inversions on the Wigner-degenerate fields are given by
\begin{alignat}{2}
P\psi_{n}(x)P^{-1} &=
\eta^{*}_{n} \gamma^{0} \psi_{n}(\mathscr{P} x) \, , 
\quad
&P \bar{\psi}_{n}(x)P^{-1} &=
\eta_{n} \bar{\psi}_{n}(\mathscr{P} x) \gamma^{0} \, , 
\label{eq:P_W_field_2} \\
T\psi_{n}(x)T^{-1} &=
e^{i n \phi}\gamma^{1}\gamma^{3}\psi_{-n}(\mathscr{T}x) \, ,
\quad
&T \bar{\psi}_{n}(x)T^{-1} &=
e^{-i n \phi} \bar{\psi}_{-n}(\mathscr{T}x) \gamma^{3}\gamma^{1} \, ,
\label{eq:T_W_field_3}
\end{alignat}
which can be expressed in the doublet form~\eqref{eq:doublet}-\eqref{eq:doublet_D_conj}
\begin{alignat}{2}
P\Psi(x)P^{-1}
&=\gamma^{0} D^{*}(\mathscr{P})
\Psi(\mathscr{P}x) \, , 
\quad 
&P \overline{\Psi}(x) P^{-1}
&= \overline{\Psi}(\mathscr{P}x) D(\mathscr{P}) \gamma^{0}
\, , 
\label{eq:P_W_field_D2} \\
T\Psi(x)T^{-1}
&=\gamma^{1}\gamma^{3} D(\mathscr{T}) \Psi(\mathscr{T}x) \, ,
\quad 
&T \overline{\Psi}(x) T^{-1}
&= \overline{\Psi}(\mathscr{T}x) D(\mathscr{T}) \gamma^{3}\gamma^{1}
\, .
\label{eq:T_W_field_D2}
\end{alignat}
The general formula of the two discrete inversions on the Wigner doublets~\eqref{eq:P_state_W_ab}-\eqref{eq:T_state_W_ab} can be simplified via Eqs.~\eqref{mat:P_state_W_ab_2},\eqref{mat:T_state_W_ab_2}, and \eqref{eq:PT_W_phi_2}
\begin{align}
    P|\p,\sigma,n; a\rangle&= \eta_{n} |-\p,\sigma,n; a\rangle \, ,
    \label{eq:P_state_W_a_2} \\
    P|\p,\sigma,n; a^{c}\rangle&= -\eta_{n}^{*} |-\p,\sigma,n; a^{c}\rangle \, ,
    \label{eq:P_state_W_b_2}\\
    T|\p,\sigma,n; a\rangle&=(-1)^{\frac{1}{2}-\sigma} e^{-in\phi} |-\p,-\sigma,-n; a\rangle \, ,
    \label{eq:T_state_W_a_2} \\
    T|\p,\sigma,n; a^{c}\rangle&=(-1)^{\frac{1}{2}-\sigma} e^{in\phi} |-\p,-\sigma,-n; a^{c}\rangle \, .
    \label{eq:T_state_W_b_2}
\end{align}
It is remarkable to note that the Wigner degeneracy generally provides an additional contribution to the action of $T^{2}$ on the one-particle states with $e^{i\phi} \neq 1$
\begin{align}
T^{2}|\p,\sigma,\pm\textstyle{\frac{1}{2}}; a^{(c)}\rangle 
&= - e^{\pm i\phi^{(c)}} |\p,\sigma,\pm\textstyle{\frac{1}{2}}; a^{(c)} \rangle
= - \left[ e^{\pm i\phi} \right]^{(*)} |\p,\sigma,\pm\textstyle{\frac{1}{2}}; a^{(c)} \rangle
\, ,
\label{eq:Ut2_W}
\end{align}
which recovers the standard case of Eq.~\eqref{eq:Ut_2} with $e^{i\phi} = 1$ and we have imposed the relation of the time-reversal phases~\eqref{eq:PT_W_phi_2} in the second equality. In particular, the action of $T^{2}$ on the one-particle states will have an opposite sign to the standard case with $e^{i\phi} = -1$.

It is crucial to
understand how various bilinear forms transform under the two inversions, particularly for constructing Lagrangians.
The bilinear forms of the Wigner-degenerate fields $\bar{\chi}_{m} \Gamma \psi_{n}$ involve five basic choices of the matrix $\Gamma$:
\begin{align}
\mathds{1}, \ \gamma^{\mu}, \ \sigma^{\mu \nu} \equiv \dfrac{i}{2} \left[\gamma^{\mu}, \gamma^{\nu}  \right], \ \gamma^{\mu} \gamma^{5}, \ \gamma^{5}
\, .
\label{G_bi_1}
\end{align}
Applying the spatial parity transformation~\eqref{eq:P_W_field_2} to these bilinear forms yields:
\begin{align}
P \left[ \bar{\chi}_{m} (x) \Gamma \psi_{n} (x) \right] P^{-1} 
= \left(\tilde{\eta}_{m} / \eta_{n} \right) \bar{\chi}_{m}(\mathscr{P} x) \gamma^{0} \Gamma
 \gamma^{0} \psi_{n}(\mathscr{P} x)
\, ,
\label{eq:P_W_bi_1}
\end{align}
where $\tilde{\eta}_{m}$ and $\eta_{n}$ are intrinsic parities of $\chi_{m}$ and $\psi_{n}$ respectively.
Taking the matrix $\Gamma$ as one of the five standard forms~\eqref{G_bi_1}, the bilinear form transforms as
a scalar, vector, tensor, pseudo- (or axial-) vector, and pseudoscalar, respectively up to a ratio of intrinsic parities~\footnote{$\tilde{\eta}_{m} \eta^{*}_{n} = \tilde{\eta}_{m} / \eta_{n}$, with $\eta_{n} \eta^{*}_{n} = 1$.}. This result is consistent with CQFT since the transformation matrix of the spatial parity~\eqref{mat:P_state_W_ab_2} is taken to be diagonal with respect to the Wigner degeneracy here.
An exotic phenomenon emerges when we apply the time-reversal~\eqref{mat:T_state_W_ab_2} to the bilinear form
\begin{align}
T \left[ \bar{\chi}_{m} (x) \Gamma \psi_{n} (x) \right] T^{-1} 
= 
e^{-i (m \tilde{\phi} - n \phi)} \bar{\chi}_{-m}(\mathscr{T}x) \Gamma_t \psi_{-n}(\mathscr{T}x)
\, ,
\label{eq:T_W_bi_1}
\end{align}
with $\Gamma_t \equiv \mathcal{T} \Gamma^{*} \mathcal{T}^{-1}$, 
$\mathcal{T} \equiv i \gamma^{1}\gamma^{3} = \mathcal{T}^{\dag} = \mathcal{T}^{-1} = -\mathcal{T}^{*}$.
We see that $T$ flips the Wigner degeneracy but remains a similar formula as the standard case up to a discrepancy of the time-reversal phase factor.
However, it is important to note that there may exist an additional internal symmetry unitary operator $\hat{S}_{T}$ that acts on the Wigner doublet as
\begin{align}
\hat{S}_{T} |\p,\sigma,n; a\rangle = e^{-i n \phi} |\p,\sigma,-n; a\rangle \, , 
\quad
\hat{S}_{T} |\p,\sigma,n; a^{c}\rangle = e^{i n \phi} |\p,\sigma,-n; a^{c}\rangle \, , 
\label{eq:S_re_T_0}
\end{align}
but trivially on usual particles
\begin{align}
\hat{S}_{T} |\p,\sigma; a^{(c)}\rangle &= \eta_{S} |\p,\sigma; a^{(c)}\rangle \, , 
\ \ \text{with} \ \ 
\eta_{S} = \pm 1 \, ,
\end{align}
so that $\hat{S}_{T}=\hat{S}_{T}^{-1}=\hat{S}_{T}^{\dag}$.
Applying $\hat{S}_{T}$ to the Wigner-degenerate fields $\psi_{n}(x)$, $n=\pm\frac{1}{2}$~\eqref{eq:sf_W_20} yields
\begin{align}
\hat{S}_{T} \psi_{n}(x) \hat{S}_{T}^{-1}
= e^{i n \phi} \psi_{-n}(x) 
\, , 
\quad
\hat{S}_{T} \bar{\psi}_{n}(x) \hat{S}_{T}^{-1}
= e^{-i n \phi} \bar{\psi}_{-n}(x) 
\label{eq:S_re_T_field_1}
\end{align}
associated with its doublet form
\begin{align}
\hat{S}_{T} \Psi (x) \hat{S}_{T}^{-1} = D(\mathscr{T}) \Psi (x) \, ,
\quad
\hat{S}_{T} \overline{\Psi} (x) \hat{S}_{T}^{-1} = \overline{\Psi} (x) D(\mathscr{T}) \, . 
\label{eq:S_re_T_field_2}
\end{align}
Then, utilizing this internal symmetry operator, one can redefine the time-reversal operator as
\begin{align}
T' \equiv \hat{S}_{T}^{-1} T \, ,
\end{align}
which leads to the time-reversal of one-particle states without Wigner degeneracy mixing
\begin{align}
T' |\p,\sigma,n; a^{(c)}\rangle&=(-1)^{\frac{1}{2}-\sigma} |-\p,-\sigma,n; a^{(c)} \rangle \, , \\
T' |\p,\sigma; a^{(c)}\rangle &= (-1)^{j-\sigma} \eta_{S}^{(*)} |-\p, -\sigma; a^{(c)}\rangle \, .
\end{align}
Thus, all particles transform in the usual way under $T'$.
The redefined time-reversal transformation on the Wigner-degenerate fields $\psi_{n}(x)$, $n=\pm\frac{1}{2}$~\eqref{eq:sf_W_20} can be directly obtained by combining Eqs.~\eqref{eq:T_W_field_3} and \eqref{eq:S_re_T_field_1}:
\begin{align}
T' \psi_{n}(x) T^{\prime -1} = \gamma^{1}\gamma^{3} \psi_{n}(\mathscr{T}x) \, ,
\quad
T' \bar{\psi}_{n}(x) T^{\prime -1} = \bar{\psi}_{n}(\mathscr{T}x) \gamma^{3}\gamma^{1} \, ,
\end{align}
and its doublet form
\begin{align}
T' \Psi (x) T^{\prime -1} = \gamma^{1}\gamma^{3} \Psi (\mathscr{T}x) \, ,
\quad
T' \overline{\Psi} (x) T^{\prime -1} = \overline{\Psi} (\mathscr{T}x) \gamma^{3}\gamma^{1} \, , 
\end{align}
where the Wigner degeneracy mixing is eliminated as expected.
Within this redefinition, all particles transform under time-reversal in the conventional pattern.
One might suggest that a $CT$ invariant theory could be formulated to resemble a $T$ invariant Wigner QFT through an appropriate basis transformation. However, this transformation would necessarily involve superpositions of particle and antiparticle states carrying different $U(1)$ charges.
However, in our framework, charge superselection rules are required, meaning that not all superpositions of particle states are physically allowed. The Hilbert space of a system should split into noncoherent subspaces. 
This contrasts with some other theoretical frameworks incorporating Wigner degeneracy, such as Ref.~\cite{daSilvaBarbosa:2023xfy}.
Thus, the two-fold degeneracy under time-reversal can be physically meaningful only if no internal symmetry operator $\hat{S}_{T}$ exists that exchanges the two states within the Wigner doublet.
In other words, for a nontrivial $T$ invariant Wigner theory of time-reversal doublets, such an internal symmetry must be absent. This condition is crucial for self-consistency and resolves the ambiguity problem of space-time reflection operators, as discussed in Ref.~\cite{Lee:1966ik}. 
We will further explore the violation of the internal symmetry $\hat{S}_{T}$ in the presence of typical interactions in Section~\ref{sec:W_inter}.

\subsection{Charge-conjugation on the Wigner doublets}
\label{sec:C_W}

In the preceding sections, we have investigated the representations of the Poincar\'{e} group, the isometry group of the Minkowski spacetime. In this section, we introduce the charge-conjugation, which relates particles to their corresponding antiparticles. Unlike the Lorentz covariance, charge-conjugation is an internal symmetry, independent of the spacetime structure.
In the standard representation, the charge-conjugation operator $C$ interchanges particles and anti-particles with an additional phase 
\begin{align}
C |\p,\sigma;a\rangle &= \eta_{C} |\p,\sigma;a^{c}\rangle \, ,
\label{eq:C_state_a} \\
C |\p,\sigma;a^{c}\rangle &= \eta_{C}^{c} |\p,\sigma;a\rangle
= \eta_{C}^{*} |\p,\sigma;a\rangle \, ,
\label{eq:C_state_b} \\
C^{2} |\p,\sigma;a^{(c)}\rangle &= \eta_{C} \eta_{C}^{c} |\p,\sigma;a^{(c)}\rangle
= |\p,\sigma;a^{(c)}\rangle
\, ,
\label{eq:C_state_ab_1}
\end{align}
where the phases $\eta_{C}$ and $\eta_{C}^{c}$ are the charge-conjugation parities for the particle and its antiparticle respectively. 
The relation $\eta_{C}^{*} = \eta_{C}^{c}$ can be further derived from the causality condition, which requires the causal field to be transformed into another field commuting at space-like separations.
In Section~\ref{sec:Spin_m_1}, we observed that the time-reversal operator $T$ can exchange the particle species within the Wigner doublet framework, which does not occur in the standard representation.
Utilizing the same methodology, we consider the unitary charge-conjugation operator $C$ acting on the Wigner-degenerate particle state $|\p,\sigma,n;a\rangle$. 
Under charge-conjugation, this state transforms into a superposition of anti-particle states $|\p,\sigma,n;a^{c}\rangle$, summed over the degenerate degrees of freedom
\begin{equation}
    C |\p,\sigma,n;a\rangle=\sum_{n'}D_{n'n}(C)|\p,\sigma,n';a^{c}\rangle \, .
    \label{eq:C_state_W_ab_1}
\end{equation}
Similarly in the reverse process, the anti-particle state $|\p,\sigma,n;a^{c}\rangle$ transforms to a superposition of particle states $|\p,\sigma,n;a\rangle$
\begin{equation}
    C |\p,\sigma,n;a^{c}\rangle=\sum_{n'}D^{c}_{n'n}(C)|\p,\sigma,n';a\rangle \,.
    \label{eq:C_state_W_ab_2}
\end{equation}
Therefore, under charge-conjugation, the annihilation and creation operators transform as
\begin{alignat}{2}
    C a^{\dag}_{n}(\p,\sigma) C^{-1} &=\sum_{n'}D_{n'n}(C)a^{c\dag}_{n'}(\p,\sigma)\, ,
    \quad
    &C a_{n}(\p,\sigma) C^{-1} &=\sum_{n'}D^{*}_{n'n}(C)a^{c}_{n'}(\p,\sigma)\, ,
    \\
    C a^{c\dag}_{n}(\p,\sigma) C^{-1} &=\sum_{n'}D^{c}_{n'n}(C)a^{\dag}_{n'}(\p,\sigma)\, ,
    \quad
    &C a^{c}_{n}(\p,\sigma) C^{-1} &=\sum_{n'}D^{c*}_{n'n}(C)a_{n'}(\p,\sigma)\, ,
\end{alignat}
which induce the charge-conjugation transformation for the Wigner-degenerate fields $\psi_{n}(x)$, $n=\pm\frac{1}{2}$~\eqref{eq:sf_W_20} 
\begin{align}
    C \psi_{n}(x) C^{-1} 
    &=-i\gamma^{2}\int\frac{d^{3}p}{(2\pi)^{3}}\frac{1}{\sqrt{2E_{\p}}}\sum_{\sigma,n'}
    \biggl[e^{-ip\cdot x}v^{*}(\p,\sigma)D^{*}_{n'n}(C) a^{c}_{n'}(\p,\sigma)
    \nonumber\\    &\quad\quad\quad\quad\quad\quad\quad\quad\quad\quad\quad \ \ 
    +e^{ip\cdot x}u^{*}(\p,\sigma)D^{c}_{n'n}(C) a^{\dag}_{n'}(\p,\sigma)\biggr]
    \, ,
    \label{eq:C_psi}
\end{align}
where we have imposed the following identities compatible with the polarizations~\eqref{sf_u_W_20}-\eqref{sf_v_W_20}
\begin{align}
    -i\gamma^{2}u^{*}(\p,\sigma) = v(\p,\sigma)\,, 
    \quad
    -i\gamma^{2}v^{*}(\p,\sigma) = u(\p,\sigma)\,.
\end{align}
If we expect the charge-conjugation to map the Wigner-degenerate fields $\psi_{n}(x)$, $n=\pm\frac{1}{2}$ to a superposition of their complex conjugates $\psi^{*}_{n}(x)$~\footnote{$*$ is a shorthand for $\dag \text{T}$ so that $\psi^{*}_{n}(x)$ is a column vector with components that are the adjoints of those of $\psi_{n}(x)$.}, $n=\pm\frac{1}{2}$, then the transformation matrices $D(C)$ and $D^{c}(C)$ must satisfy the relation
\begin{equation}
D^{*}(C)=D^{c}(C) \, ,
\label{eq:C_W_psi_2}
\end{equation}
which takes a similar form as that in the standard case~\eqref{eq:C_state_b}.
Then, we obtain the action of the charge-conjugation on the Wigner-degenerate fields
\begin{align}
C \psi_{n}(x) C^{-1} &= -i\sum_{n'}D^{*}_{n'n}(C)\gamma^{2}\psi^{*}_{n'}(x)\, ,
\label{eq:C_W_field_2} \\
C \bar{\psi}_{n}(x) C^{-1} &= i\sum_{n'}D_{n'n}(C) \psi^{\text{T}}_{n'}(x) \gamma^{0} \gamma^{2}\, ,
\label{eq:C_W_field_3}
\end{align}
and its associated doublet form
\begin{align}
C \Psi(x) C^{-1}
= -i \gamma^{2} D^{\dag}(C) \Psi^{*}(x) 
\, ,
\quad
C \overline{\Psi}(x) C^{-1}
= i \Psi^{\text{T}}(x) D(C) \gamma^{0} \gamma^{2} 
\, . 
\label{eq:C_W_field_D2}
\end{align}
The matrix $D(C) \in U(2)$ can be factorized by a factor $\theta$ and a three dimensional vector $\bm \theta = (\theta_1, \theta_2, \theta_3)$
\begin{align}
D(C(\theta, \bm \theta)) 
= e^{i \frac{\theta}{2}}\exp \left( i \theta_{a} \tau^{a} \right) 
\, ,
\end{align}
with $\tau^{a}=\frac{\sigma^{a}}{2}$, $a=1,2,3$.
The explicit matrix form of $D(C(\theta, \bm \theta))$ is then given by
\begin{align}
D(C(\theta, \bm \theta)) &= e^{i \frac{\theta}{2}}
\left[\begin{matrix}
\cos \frac{|\bm \theta|}{2} + i \hat{\theta}_{3} \sin \frac{|\bm \theta|}{2} & (\hat{\theta}_{2} + i \hat{\theta}_{1}) \sin \frac{|\bm \theta|}{2} \\
(-\hat{\theta}_{2} + i \hat{\theta}_{1}) \sin \frac{|\bm \theta|}{2} & \cos \frac{|\bm \theta|}{2} - i \hat{\theta}_{3} \sin \frac{|\bm \theta|}{2}
\end{matrix}\right]
\, ,
\label{mat:C_state_W_ab_01} 
\end{align}
with $\hat{\theta}_{a} \equiv \theta_{a}/|\bm \theta|$, $a=1,2,3$~\footnote{If $\theta_1 = \theta_2 = \theta_3 = 0$, we adopt the convention $\hat{\theta}_{1} = \hat{\theta}_{2} = \hat{\theta}_{3} \equiv 1$.}, $|\bm \theta| \equiv \sqrt{\theta_{1}^2+ \theta_{2}^2+ \theta_{3}^2}$.

Applying the charge-conjugation~\eqref{eq:C_W_field_2}-\eqref{eq:C_W_field_3} to the Wigner-degenerate bilinear forms yields:
\begin{align}
C  \left[ \bar{\chi}_{m} (x) \Gamma \psi_{n} (x) \right]  C^{-1}  
&= \sum_{m',n'} \tilde{D}_{m'm}(C) D^{*}_{n'n}(C)
\left[ \bar{\psi}_{n'}(x) \Gamma_c \chi_{m'}(x) \right]
\nonumber \\
&= \sum_{m',n'} \tilde{D}_{m'm}(C) D^{*}_{n'n}(C)
\left[ \bar{\chi}_{m'}(x) \gamma^{0} \Gamma_c^{\dag} \gamma^{0} \psi_{n'}(x) 
\right]^{\dag}
\, ,
\label{eq:C_W_bi_1}
\end{align}
where we have inserted anticommuting relations for two fermionic spinors and ignored a c-number anticommutator.
We define $\Gamma_c \equiv \left( \mathcal{C}^{-1} \Gamma \mathcal{C} \right)^{\text{T}} = \mathcal{C}^{-1} \Gamma^{\text{T}} \mathcal{C}$
with $\mathcal{C} \equiv -i \gamma^{2} \gamma^{0} = - \mathcal{C}^{-1} = - \mathcal{C}^{\dag} = - \mathcal{C}^{\text{T}}$, satisfying $\mathcal{C}^{-1} \gamma^{\mu} \mathcal{C} = - \gamma^{\mu T}$. 
We notice that the expression inside the bracket of Eq.~\eqref{eq:C_W_bi_1} has a similar form to the standard representation in CQFT. 
If the transformation matrices $\tilde{D}(C)$ and $D(C)$ for Wiger-degenerate the fields $\chi_{m} (x)$ and $\psi_{n}(x)$ are diagonal, no Wigner degeneracy mixing will emerge, and Eq.~\eqref{eq:C_W_bi_1} will reduce to the standard case.
In contrast, the non-diagonal $\tilde{D}(C)$ and $D(C)$ introduce the Wigner degeneracy mixing and explicitly break the charge-conjugation symmetry, unless some particular combination of the Wigner-degenerate fields is imposed to restore the charge-conjugation symmetry.

\section{A challenge to the $CPT$ theorem}
\label{sec:cpt_thm}

A fundamental property of local QFT, first established by Pauli, Zumino, and Schwinger states that, in any case, $CPT$ remains an invariance of the theory -- the celebrated $CPT$ theorem, which arises directly from Lorentz invariance in the standard representation.
A remarkable consequence of this theorem is that the $S$-matrix for an arbitrary process is directly related to the $S$-matrix for the inverse process, where all spins are reversed, and particles are replaced by their corresponding antiparticles, and vice versa.
However, the presence of Wigner degeneracy mixing in $CPT$ transformations introduces additional complexity, so a more careful and thorough analysis of this generic invariance is necessary.
In this section, we investigate the $CPT$ transformation on the Wigner-degenerate states and spinor fields. 
These results will serve as a foundation for studying interacting theories, which we will explore further in Section~\ref{sec:W_inter}.

The combined action of general $P$~\eqref{eq:P_state_W_ab}, $T$~\eqref{eq:T_state_W_ab}, and $C$~\eqref{eq:C_state_W_ab_1}-\eqref{eq:C_state_W_ab_2} transformations leads to the $CPT$ transformation on Wigner-degenerate particle and antiparticle states
\begin{align}
    \Theta|\p,\sigma,n;a\rangle&=(-1)^{1/2-\sigma}\sum_{n'} \Xi_{n'n}|\p,-\sigma,n';a^{c}\rangle\,,\label{eq:CPT_part_W_1}\\
    \Theta|\p,\sigma,n;a^{c}\rangle&=(-1)^{1/2-\sigma}\sum_{n'} \Xi^{c}_{n'n}|\p,-\sigma,n';a\rangle\,,\label{eq:CPT_part_W_2}
\end{align}
where $\Theta \equiv C P T$ is an antiunitary and antilinear operator. The associated unitary matrices
\begin{align}
\Xi^{(c)} = D^{(c)}(C)D^{(c)}(\mathscr{P})D^{(c)}(\mathscr{T}) \, ,
\label{eq:CPT_Xi_1}
\end{align}
generally induce mixing of Wigner degeneracies under $CPT$ unless they are diagonal.
Imposing properties of $C$, $P$, and $T$ matrices in Eqs.~\eqref{eq:PT_W_phi_2} and \eqref{eq:C_W_psi_2} further yields the relation
\begin{align}
\Xi^{*} = - \Xi^{c} \, .
\label{eq:CPT_Xi_Xi}
\end{align}
The explicit formula of $\Xi$~\eqref{eq:CPT_Xi_1} can be obtained by directly multiplying Eqs.~\eqref{mat:P_state_W_ab_2}, \eqref{mat:T_state_W_ab_2}, and \eqref{mat:C_state_W_ab_01}
\begin{align}
\Xi_{n'n} &= \eta_{-n} e^{-in\phi} D_{n' -n}(C)
\nonumber \\
&= e^{i\frac{\theta}{2}}
    \left[\begin{matrix}
        \eta_{-\frac{1}{2}}(i\hat{\theta}_{1}+\hat{\theta}_{2})e^{-i\frac{\phi}{2}} \sin \frac{|\boldsymbol{\theta}|}{2} 
        & \eta_{+\frac{1}{2}}\left(\cos \frac{|\boldsymbol{\theta}|}{2}+i\hat{\theta}_{3}\sin \frac{|\boldsymbol{\theta}|}{2}\right) e^{i\frac{\phi}{2}} \\
        \eta_{-\frac{1}{2}}\left(\cos \frac{|\boldsymbol{\theta}|}{2}-i\hat{\theta}_{3}\sin \frac{|\boldsymbol{\theta}|}{2}\right) e^{-i\frac{\phi}{2}} 
        & \eta_{+\frac{1}{2}}(i\hat{\theta}_{1}-\hat{\theta}_{2})e^{i\frac{\phi}{2}}\sin \frac{|\boldsymbol{\theta}|}{2}
    \end{matrix}\right]\,.
\label{eq:cpt_matrix}
\end{align}
In CQFT without the Wigner degeneracy, $(CPT)^{2}$ transformation on the states can be calculated by Eqs.~\eqref{eq:Up}, \eqref{eq:Ut}, and \eqref{eq:C_state_a}-\eqref{eq:C_state_ab_1}
\begin{align}
    \Theta^{2}|\p,\sigma;a\rangle
    = (-)^{2 \times \frac{1}{2}} \eta_{C}^{*} \eta_{P}^{*} \eta_{T}^{*} \eta_{C}^{c} \eta_{P}^{c} \eta_{T}^{c} |\p,\sigma;a\rangle
    = |\p,\sigma;a\rangle
    \, ,
    \label{CPT2_0}
\end{align}
where $\eta_{P}^{c} = - \eta_{P}^{*}$, $\eta_{T}^{c} = \eta_{T}^{*}$, $\eta_{C}^{c} = \eta_{C}^{*}$ and we conventionally set $\eta_{C} \eta_{P} \eta_{T} = 1$ for particle $a$~\footnote{In some literature~\cite{Langacker:2017uah}, intrinsic phases are chosen such that $\Theta^{2} = (-)^{2j} \boldsymbol{1}$, where $j$ is the particle spin.}. Thus, the one-particle state is physically invariant under $(CPT)^{2}$.
In contrast, Eqs.~\eqref{eq:CPT_part_W_1}-\eqref{eq:CPT_Xi_Xi} induce $(CPT)^{2}$ on the Wigner doublets
\begin{align}
\Theta^{2}|\p,\sigma,n;a\rangle&=\sum_{n'}\Xi^{*2}_{n'n}|\p,\sigma,n';a\rangle\,,
\label{eq:Theta1}\\
\Theta^{2}|\p,\sigma,n;a^{c}\rangle&=\sum_{n'}\Xi^{2}_{n'n}|\p,\sigma,n';a^{c}\rangle\,,
\label{eq:Theta2}
\end{align}
which suggests a possible change of particle type due to Wigner degeneracy mixing.
We will explore several specific $CPT$ configurations in Sections~\ref{sec:CPT_d_W} and~\ref{sec:CPT_bd_W}. 
Although our discussion so far has focused on operators that create and annihilate particles in free-particle states, this formalism can be extended naturally to `in' and `out' states.
The $CPT$ transformation of free Wigner-degenerate fields can be obtained directly from three discrete inversions given in Eqs.~\eqref{eq:P_W_field_2}-\eqref{eq:T_W_field_3} and \eqref{eq:C_W_field_2}-\eqref{eq:C_W_field_3}
\begin{align}
\Theta \psi_{n}(x) \Theta^{-1} 
&= \sum_{n'} \Xi_{n'n}^{*} \gamma^{5} \psi_{n'}^{*}(-x) 
\, ,
\label{eq:CPT_W_field_1} \\
\Theta  \bar{\psi}_{n}(x) \Theta ^{-1} &= \sum_{n'} \Xi_{n'n} \psi_{n'}^{\text{T}} (-x) \gamma^{5} \gamma^{0}
\, ,
\label{eq:CPT_W_field_2}
\end{align}
associated with the doublet formalism 
\begin{align}
\Theta \Psi(x) \Theta^{-1}
= \gamma^{5} \Xi^{\dag} \Psi^{*}(-x) 
\, ,
\quad
\Theta \overline{\Psi}(x) \Theta^{-1}
= \Psi^{\text{T}}(-x) \Xi \gamma^{5} \gamma^{0} 
\, . 
\label{eq:CPT_W_field_D2}
\end{align}
Applying Eqs.~\eqref{eq:CPT_W_field_1}-\eqref{eq:CPT_W_field_2} to the $CPT$ transformation on general bilinears of the Wigner-degenerate fields gives
\begin{align}
\Theta \left[ \bar{\chi}_{m} (x) \Gamma \psi_{n} (x) \right] \Theta^{-1}
&= (-1)^{r} \sum_{m',n'} \tilde{\Xi}_{m'm} \Xi_{n'n}^{*}
\left[ \bar{\chi}_{m'} (-x) \Gamma \psi_{n'}(-x) \right]^{\dag}
\, ,
\label{eq:CPT_W_bi_1}
\end{align}
with
\begin{align}
\chi_{m'}^{\text{T}} (-x) \gamma^{5} \gamma^{0} 
\Gamma^{*}
\gamma^{5} \psi_{n'}^{*}(-x)
&= \left[ \bar{\chi}_{m'} (-x) \left( \gamma^{5} 
\Gamma
\gamma^{5} \right) \psi_{n'}(-x) \right]^{\dag}
\, ,
\\
\gamma^{5} \Gamma \gamma^{5} &= (-1)^{r} \Gamma \, ,
\end{align}
where $\chi_{m}$ and $\psi_{n}$ are anticommuting fields as in Eq.~\eqref{eq:C_W_bi_1}. 
$\Gamma$ is a product of $r$ Dirac matrices. 
We notice that the item in the bracket of Eq.~\eqref{eq:CPT_W_bi_1} is similar to that in CQFT. 
However, non-diagonal $\tilde{\Xi}$ and $\Xi$ may lead to nontrivial Wigner degeneracy mixing, which can explicitly break the $CPT$ symmetry, unless a specific combination of Wigner-degenerate fields is deliberately chosen to restore it. 
In other words, this mixing of the Wigner degeneracies would not change the Lagrangian only if there is a superposition over all Wigner indices. Therefore, the $CPT$ invariant Lagrangian should be constructed via Wigner doublets. A detailed discussion on this construction will be presented in Section~\ref{sec:can_W_1}.
Additionally, this feature will play a crucial role in constructing interaction terms in Section~\ref{sec:W_inter}.

In particular, if one chooses the same $CPT$ matrix for all Wigner-degenerate fields (even for different particle species)~\footnote{This convention is analogous to that in CQFT.}, inserting $\tilde{\Xi} = \Xi$ to Eq.~\eqref{eq:CPT_W_bi_1} gives
\begin{align}
\Theta \left[ \sum_{n} \bar{\chi}_{n} (x) \Gamma \psi_{n} (x) \right] \Theta^{-1}
= (-1)^{r} \left[\sum_{n} \bar{\chi}_{n} (-x) \Gamma \psi_{n}(-x) \right]^{\dag}
\, .
\label{eq:CPT_W_bi_inner_0}
\end{align}
Although the general $\Xi$~\eqref{eq:cpt_matrix} depends on eight parameters, the $CPT$ transformation can be classified into two distinct classes. 
Since $\Theta$ is antilinear and antiunitary, the $2 \times 2$ unitary matrix $\Xi$ can always be transformed into either a diagonal or an anti-diagonal form by appropriately redefining the basis states.
In the anti-diagonal case, the diagonal elements vanish, while the off-diagonal elements (phases) are complex conjugates of each other, assuming no additional constraints are imposed, analogous to the time-reversal transformation discussed in Section~\ref{sec:PT_1} and Appendix~\ref{sec:T_block_1}.
To clearly illustrate the nontrivial effects of $\Xi$, we simplify its structure by choosing the suitable parameters, ensuring that $\Xi$ takes either a diagonal form (see Section~\ref{sec:CPT_d_W}) or an anti-diagonal form (see Section~\ref{sec:CPT_bd_W}).

\subsection{Diagonal $CPT$}
\label{sec:CPT_d_W}


Let us first consider the diagonal configuration of the $CPT$ matrix $\Xi$~\eqref{eq:cpt_matrix}, factorized as
\begin{align}
    \Xi
    =
    \left[\begin{matrix}
        e^{i\frac{\vartheta_{+}}{2}} & 0 \\
        0 & e^{i\frac{\vartheta_{-}}{2}}
    \end{matrix}\right]
    \equiv 
    e^{i\frac{\theta}{2}} \left[\begin{matrix}
        \eta_{-\frac{1}{2}}(i\hat{\theta}_{1}+\hat{\theta}_{2})e^{-i\frac{\phi}{2}} & 0 \\
        0 & \eta_{+\frac{1}{2}}(i\hat{\theta}_{1}-\hat{\theta}_{2})e^{i\frac{\phi}{2}}
    \end{matrix}\right]
    \, ,
    \label{eq:cpt_matrix1}
\end{align}
with $\vartheta_{\pm}\in\mathbb{R}, \ \hat{\theta}_{3}=0, \ |\boldsymbol{\theta}|=\pi$, i.e., diagonal elements of $D(C)$~\eqref{mat:C_state_W_ab_01} vanish. Using Eqs.~\eqref{eq:CPT_part_W_1}-\eqref{eq:CPT_part_W_2}, we obtain
\begin{align}
    \Theta |\p,\sigma,n;a\rangle&=(-1)^{1/2-\sigma}e^{i\frac{\vartheta_{2n}}{2}}|\p,-\sigma,n;a^{c}\rangle\,,\\
     \Theta |\p,\sigma,n;a^{c}\rangle&=-(-1)^{1/2-\sigma}e^{-i\frac{\vartheta_{2n}}{2}}|\p,-\sigma,n;a\rangle\,,
     \label{eq:CPT_W_part_2d}
\end{align}
and Eqs.~\eqref{eq:Theta1}-\eqref{eq:Theta2} give
\begin{align}
    \Theta^{2}|\p,\sigma,n;a\rangle
    =e^{-i\vartheta_{2n}}|\p,\sigma,n;a\rangle
    \, ,
    \quad
    \Theta^{2}|\p,\sigma,n;a^{c}\rangle
    =e^{i\vartheta_{2n}}|\p,\sigma,n;a^{c}\rangle
    \, .
\end{align}
Note that $e^{i\vartheta_{2n}}=-1$, with $n= +\frac{1}{2} \ \text{or} \ -\frac{1}{2}$, provides the new possibility of an additional sign relative to the result in CQFT~\eqref{CPT2_0}. 
Furthermore, if we require the action of $\Theta^{2}$ on the particle and anti-particle states corresponding to any Wigner index to be the same, the phases in Eq.~\eqref{eq:cpt_matrix1} must be set to $e^{i\vartheta_{+}}=e^{i\vartheta_{-}}=\pm 1$ which then corresponds to $\Theta^{2}=\pm \boldsymbol{1}$ respectively.
The general results of $CPT$ on Wigner-degenerate fields~\eqref{eq:CPT_W_field_1}-\eqref{eq:CPT_W_field_2} can be simplified in the diagonal configuration of Eq.~\eqref{eq:cpt_matrix1}
\begin{align}
\Theta \psi_{n}(x) \Theta^{-1} = 
e^{-i\frac{\vartheta_{2n}}{2}} \gamma^{5} \psi_{n}^{*}(-x) 
\, ,
\quad
\Theta  \bar{\psi}_{n}(x) \Theta ^{-1} = 
e^{i\frac{\vartheta_{2n}}{2}} \psi_{n}^{\text{T}} (-x) \gamma^{5} \gamma^{0}
\, ,
\label{eq:CPT_W_field_12d}
\end{align}
with the corresponding doublet form
\begin{align}
\Theta\Psi(x)\Theta^{-1} = \Xi^{*} \gamma^{5} \Psi^{*}(-x)
\, , \quad
\Theta\overline{\Psi}(x)\Theta^{-1} =\Psi^{\text{T}}(-x) \gamma^{5} \gamma^{0} \Xi 
\, . \label{eq:cpt_psi_1}
\end{align}
Then, the general result of the $CPT$ on bilinears~\eqref{eq:CPT_W_bi_1} also further simplifies to
\begin{align}
\Theta \left[ \bar{\chi}_{m} (x) \Gamma \psi_{n} (x) \right] \Theta^{-1}
&= (-1)^{r} e^{i\frac{\tilde{\vartheta}_{2m}}{2}} e^{-i\frac{\vartheta_{2n}}{2}}
\left[ \bar{\chi}_{m} (-x) \Gamma \psi_{n}(-x) \right]^{\dag}
\, ,
\label{CPT_W_bi_d_1}
\end{align}
which remains the same structure as CQFT up to a phase factor and does not introduce Wigner degeneracy mixing. 
Thus, under the diagonal configuration of the $CPT$ matrix~\eqref{eq:cpt_matrix}, one can always choose intrinsic $CPT$ phases so that every Poincar\'{e} invariant term in the Lagrangian will be mapped to its Hermitian
conjugate evaluated at $-x$, leaving the action invariant.
This result aligns with our original expectation since the diagonal $\Xi$~\eqref{eq:cpt_matrix}
does not introduce any mixing of the Wigner degeneracy.

\subsection{Anti-diagonal $CPT$}
\label{sec:CPT_bd_W}

Next, we consider the anti-diagonal configuration of the $CPT$ matrix $\Xi$~\eqref{eq:cpt_matrix}
\begin{align}
\Xi
= \left[\begin{matrix}
        0 & e^{i\frac{\varphi}{2}} \\
        e^{-i\frac{\varphi}{2}} & 0
    \end{matrix}\right]
\equiv
    \left[\begin{matrix}0 & \eta_{+\frac{1}{2}} e^{i\frac{\theta_3}{2}} e^{i\frac{\phi}{2}} \\
        \eta^{*}_{+\frac{1}{2}} e^{-i\frac{\theta_3}{2}} e^{-i\frac{\phi}{2}} & 0
    \end{matrix}\right]
    \,,
\label{eq:cpt_matrix2}
\end{align}
with $\varphi\in\mathbb{R}, \ \eta_{+\frac{1}{2}}=\eta^{*}_{-\frac{1}{2}}$, $(\theta, \boldsymbol{\theta})=(0, 0,0, \theta_3)$. That is, $D(C)$~\eqref{mat:C_state_W_ab_01} becomes a diagonal element of the $SU(2)$ group. 
Using Eqs.~\eqref{eq:CPT_part_W_1}-\eqref{eq:CPT_part_W_2}, we find
\begin{align}
    \Theta |\p,\sigma,n;a\rangle&=(-1)^{1/2-\sigma}e^{-i n \varphi}|\p,-\sigma,-n;a^{c}\rangle\,,\\
     \Theta |\p,\sigma,n;a^{c}\rangle&=-(-1)^{1/2-\sigma}e^{i n \varphi}|\p,-\sigma,-n;a\rangle\,.
\end{align}
Imposing the general formula of Eqs.~\eqref{eq:Theta1}-\eqref{eq:Theta2}, the result of two successive $CPT$ transformations are
\begin{align}
\Theta^{2}|\p,\sigma,n;a^{(c)}\rangle&=|\p,\sigma,n;a^{(c)}\rangle
\,,
\end{align}
which agrees with the result in CQFT~\eqref{CPT2_0}.
When the anti-diagonal $\Xi$ is given by Eq.~\eqref{eq:cpt_matrix1}, the $CPT$ transformations on the Wigner-degenerate fields in Eqs.~\eqref{eq:CPT_W_field_1}-\eqref{eq:CPT_W_field_2} can be simplified to
\begin{align}
\Theta \psi_{n}(x) \Theta^{-1}  
= e^{i n \varphi} \gamma^{5} \psi_{-n}^{*}(-x)
\, ,
\quad
\Theta \bar{\psi}_{n}(x) \Theta ^{-1} 
= e^{-i n \varphi} \psi_{-n}^{\text{T}} (-x) \gamma^{5} \gamma^{0}
\, ,
\label{CPT_W_field_12bd}
\end{align}
and 
\begin{align}
\Theta\Psi(x)\Theta^{-1} = \gamma^{5} \Xi \Psi^{*}(-x)
\, , \quad
\Theta\overline{\Psi}(x)\Theta^{-1} = \Psi^{\text{T}}(-x) \Xi \gamma^{5} \gamma^{0} 
\, . \label{eq:cpt_psi_2}
\end{align}
The general result of the $CPT$ on bilinear forms of Eq.~\eqref{eq:CPT_W_bi_1} can be further simplified with the anti-diagonal $\tilde{\Xi}$ and $\Xi$ in the form of Eq.~\eqref{eq:cpt_matrix2}:
\begin{align}
\Theta \left[ \bar{\chi}_{m} (x) \Gamma \psi_{n} (x) \right] \Theta^{-1}
&= (-1)^{r} e^{-im \tilde{\varphi}} e^{in \varphi}
\left[ \bar{\chi}_{-m} (-x) \Gamma \psi_{-n}(-x) \right]^{\dag}
\, ,
\label{CPT_W_bi_bd_1}
\end{align}
which flips the Wigner degeneracies, distinguishing from the diagonal case of Eq.~\eqref{CPT_W_bi_d_1}.

\section{Canonical formalism: the Wigner doublet Lagrangian}
\label{sec:can_W_1}


In modern QFT, the canonical formalism, based on postulating the Lagrangian and applying canonical quantization, serves as the foundational starting point for analyzing any given system. The Lagrangian formalism provides a clear path to identifying symmetries such as Lorentz (or Poincar\'{e}) invariance, as well as other imposed symmetries. In this section, we will develop a suitable Lagrangian form and canonical quantization relations to describe free Wigner-degenerate spinor fields, which arise from Lorentz invariance and discrete inversions, as outlined in previous sections.

\subsection{Lagrangian of the free Wigner-degenerate spinor fields}
\label{sec:L_W2_free}

The free Lagrangian density of the massive Wigner doublet can be constructed as
\begin{equation}
    \mathcal{L}_{0}(x)=\overline{\Psi}(x)(i\gamma^{\mu}\partial_{\mu}-m)\Psi(x)
    \, , 
    \label{eq:LPsi_1}
\end{equation}
where $\Psi(x)$ is the Wigner doublet~\eqref{eq:doublet} and
$\overline{\Psi}(x)$ represents its Dirac conjugate~\eqref{eq:doublet_D_conj}. 
The parameter $m$ is their common mass.
Alternatively, this compact form can be rewritten explicitly in terms of the two-fold Wigner spinor fields $\psi_{n}$, $n=\pm\frac{1}{2}$ as
\begin{equation}
    \mathcal{L}_{0}(x)= \sum_{n} \bar{\psi}_{n}(x)(i\gamma^{\mu}\partial_{\mu}-m) \psi_{n}(x)
    \, , 
\label{eq:LPsi_2}
\end{equation}
which manifests as a direct sum of two independent free Dirac fields, each corresponding to a distinct Wigner degeneracy. It is important that there are no interaction terms coupling the two Wigner degeneracies, implying that the Euler-Lagrange equations reduce to two decoupled Dirac equations, one for each Wigner degeneracy $\psi_{n}(x)$, $n=\pm\frac{1}{2}$ respectively. 
The canonical anti-commutation relations of the Wigner-degenerate fields $\psi_{n, \ell}(x)$ and their conjugate momenta $i\psi^{\dag}_{n, \ell}(x)$ can be directly extended as
\begin{align}
\bigl\{\psi_{n, \ell}(t, \x), i\psi^{\dag}_{n', \ell'}(t, \x')  \bigr\} &= i\delta^{(3)}(\x - \x') \delta_{nn'} \delta_{\ell \ell'} \, ,
\label{eq:can_W_field_1} \\
\bigl\{\psi_{n, \ell}(t, \x), \psi_{n', \ell'}(t, \x')  \bigr\} &= 
\bigl\{\psi^{\dag}_{n, \ell}(t, \x), \psi^{\dag}_{n', \ell'}(t, \x')  \bigr\} = 0 \, ,
\label{eq:can_W_field_2}
\end{align}
which closely recover the standard free Dirac theory but include an additional Kronecker delta function $\delta_{nn'}$ due to the Wigner degeneracy. 
This ensures that the two Wigner-degenerate fields $\psi_{n}(x)$, $n=\pm\frac{1}{2}$ realize the causality condition individually.
Each $\psi_{n}(x)$ can be expanded in momentum space using the annihilation $a_{n}(\p,\sigma)$ and creation $a^{c\dag}_{n}(\p,\sigma)$ operators, which correspond to Wigner-degenerate particles and their associated antiparticles with identical mass $m$, as previously introduced in Eq.~\eqref{eq:sf_W_20}.
The canonical quantization of Eqs.~\eqref{eq:can_W_ab_1}-\eqref{eq:can_W_ab_2} remain applicable to these operators, generating one-particle states~\eqref{eq:deg_op_ab}-\eqref{eq:norm_1p_W_2}, with the particular vacuum~\eqref{eq:vac_W_ab_1}.
In our convention, both spinor field solutions $\psi_{n}(x)$, $n=\pm\frac{1}{2}$ of Eq.~\eqref{eq:sf_W_20} share the same polarizations $u(\p,\sigma)$ and $v(\p,\sigma)$ as defined in Eqs.~\eqref{sf_u_W_20}-\eqref{sf_v_W_20}.

Although the Lagrangian formalism simplifies the construction of Lorentz-invariant and symmetric theories, the calculation of the S-matrix requires an explicit expression for the interaction Hamiltonian. In general, the Hamiltonian is obtained via the Legendre transformation. It is straightforward to show that the Lagrangian density of Eq.~\eqref{eq:LPsi_1}
yields the Hamiltonian 
\begin{align}
    H_{0} &= \sum_{n} \int d^3x \left[i \psi^{\dag}_{n}(x) \dot{\psi}_{n}(x) - \mathcal{L}_{0}(x) \right]
    \nonumber \\
    &= \sum_{n} \int d^3x \ \bar{\psi}_{n}(x)(-i\gamma^{i}\partial_{i}+m) \psi_{n}(x)
    \nonumber \\
    &= \int d^3x \ \overline{\Psi}(x)(-i\gamma^{i}\partial_{i}+m)\Psi(x)
    \, ,
\label{eq:HPsi_1}
\end{align}
which expresses the Hamiltonian in terms of the Wigner-degenerate spinor fields and their conjugate momenta, as well as in the doublet form.
Applying the Fourier decomposition of the two spinor fields $\psi_{n}(x)$, $n=\pm\frac{1}{2}$~\eqref{eq:sf_W_20}, along with the ortho-normalization relations of the polarizations~\eqref{norm_uv_1}-\eqref{norm_uv_2},
we find that
\begin{align}
    H_{0} &= \int\frac{d^{3}p}{(2\pi)^{3}} \sum_{n,\sigma} E_{\p} \left[a^{\dag}_{n}(\p,\sigma) a_{n}(\p,\sigma)+a^{c\dag}_{n}(\p,\sigma) a^{c}_{n}(\p,\sigma)\right]
    \, ,
\label{eq:HPsi_2}
\end{align}
which correctly reproduces the expected free Hamiltonian. We have omitted an infinite zero-point energy shift, which is typically irrelevant for physical calculations currently.
In principle, a well-defined free-particle Lagrangian must yield a Hamiltonian that can be expressed in terms of ladder operators (up to a constant term) while ensuring a spectrum that is bounded from below. If this condition is not met, the given Lagrangian would be considered physically inconsistent.

\subsection{$P$, $C$, $CP$, $T$, and $CPT$ in the free Wigner theory}
\label{sec:symmetry_W_free}

In CQFT, the free fermion sector is individually invariant under $P$, $C$, $CP$, $T$, and $CPT$ transformations. 
However, in the presence of multiple Wigner degeneracies, these transformations may mix different Wigner degeneracies. It is therefore necessary to carefully examine how each of these transformations acts on the free Lagrangian of Wigner doublets given in Eqs.~\eqref{eq:LPsi_1}-\eqref{eq:LPsi_2}. Applying Eqs.~\eqref{eq:P_W_bi_1}, \eqref{eq:T_W_bi_1}, \eqref{eq:C_W_bi_1}, and \eqref{eq:CPT_W_bi_inner_0} with $\Gamma = \gamma^{\mu}$, 
as well as \eqref{eq:S_re_T_field_1}, we obtain
\begin{align}
P \left[ \bar{\psi}_{n} (x) \gamma^{\mu} \psi_{n} (x) \right] P^{-1} 
&= 
\left \{
\begin{array}{r c l}
&+& \bar{\psi}_{n} \gamma^{0} \psi_{n} (\mathscr{P}x) \, ,
\\ \vspace{-0.2cm} \\
&-& \bar{\psi}_{n} \gamma^{i} \psi_{n} (\mathscr{P}x) \, ,
\end{array}
\right.
\label{eq:P_W_Lag_kin_1} \\
T \left[ \bar{\psi}_{n} (x) \gamma^{\mu} \psi_{n} (x) \right] T^{-1} 
&= 
\left \{
\begin{array}{r c l}
&+& \bar{\psi}_{-n} \gamma^{0} \psi_{-n} (\mathscr{T}x) \, ,
\\ \vspace{-0.2cm} \\
&-& \bar{\psi}_{-n} \gamma^{i} \psi_{-n} (\mathscr{T}x) \, ,
\end{array}
\right.
\label{eq:T_W_Lag_kin_1} \\
\hat{S}_{T} \left[ \bar{\psi}_{n} (x) \gamma^{\mu} \psi_{n} (x) \right] \hat{S}_{T}^{-1} 
&= \bar{\psi}_{-n} (x) \gamma^{\mu} \psi_{-n} (x) \, ,
\label{eq:S_re_T_W_Lag_kin_1} \\
C  \left[ \sum_{n} \bar{\psi}_{n} (x) \gamma^{\mu} \psi_{n} (x) \right]  C^{-1}  
&= - \sum_{n} \bar{\psi}_{n} \gamma^{\mu} \psi_{n} (x)
\, ,
\label{eq:C_W_Lag_kin_1} \\
\Theta \left[ \sum_{n} \bar{\psi}_{n} (x) \gamma^{\mu} \psi_{n} (x) \right] \Theta^{-1}
&= - \sum_{n} \bar{\psi}_{n} \gamma^{\mu} \psi_{n} (-x)
\, ,
\label{eq:CPT_W_Lag_kin_1} 
\end{align}
which lead to the invariance of the kinematic terms~\eqref{eq:LPsi_1}-\eqref{eq:LPsi_2}
\begin{align}
P \left[i\overline{\Psi}(x) \gamma^{\mu} \partial_{\mu} \Psi(x) \right] P^{-1} 
&=i\overline{\Psi} \gamma^{\mu} \partial_{\mu} \Psi (\mathscr{P}x) \, ,
\label{eq:P_W_Lag_kin_2} \\
C  \left[i\overline{\Psi}(x) \gamma^{\mu} \partial_{\mu} \Psi(x) \right]  C^{-1}  
&= -i \left(\partial_{\mu} \overline{\Psi} \right) \gamma^{\mu} \Psi (x)
\, ,
\label{eq:C_W_Lag_kin_2} \\
C  P \left[i\overline{\Psi}(x) \gamma^{\mu} \partial_{\mu} \Psi(x) \right] P^{-1}  C^{-1}  
&= -i \left(\partial_{\mu} \overline{\Psi} \right) \gamma^{\mu} \Psi (\mathscr{P}x)
\, ,
\label{eq:CP_W_Lag_kin_2} \\
T \left[i\overline{\Psi}(x) \gamma^{\mu} \partial_{\mu} \Psi(x) \right] T^{-1} 
&=i\overline{\Psi} \gamma^{\mu} \partial_{\mu} \Psi (\mathscr{T}x) \, ,
\label{eq:T_W_Lag_kin_2} \\
\hat{S}_{T} \left[i\overline{\Psi}(x) \gamma^{\mu} \partial_{\mu} \Psi(x) \right] \hat{S}_{T}^{-1} 
&= i\overline{\Psi}(x) \gamma^{\mu} \partial_{\mu} \Psi(x) \, ,
\label{eq:S_re_T_W_Lag_kin_2} \\
\Theta \left[i\overline{\Psi}(x) \gamma^{\mu} \partial_{\mu} \Psi(x) \right] \Theta^{-1}
&= -i \left(\partial_{\mu} \overline{\Psi} \right) \gamma^{\mu} \Psi (-x)
\, .
\label{eq:CPT_W_Lag_kin_2} 
\end{align}
The transformations of the scalar bilinears are given by Eqs.~\eqref{eq:P_W_bi_1}, \eqref{eq:T_W_bi_1}, \eqref{eq:C_W_bi_1}, and \eqref{eq:CPT_W_bi_inner_0} with $\Gamma = \mathds{1}$ as well as \eqref{eq:S_re_T_field_1}:
\begin{align}
P \left[ \bar{\psi}_{n} (x) \psi_{n} (x) \right] P^{-1} 
&= \bar{\psi}_{n} \psi_{n} (\mathscr{P}x) \, ,
\label{eq:P_W_Lag_m_1} \\
T \left[ \bar{\psi}_{n} (x) \psi_{n} (x) \right] T^{-1} 
&= \bar{\psi}_{-n} \psi_{-n} (\mathscr{T}x) \, ,
\label{eq:T_W_Lag_m_1} \\
\hat{S}_{T} \left[ \bar{\psi}_{n} (x) \psi_{n} (x) \right] \hat{S}_{T}^{-1} 
&= \bar{\psi}_{-n} (x) \psi_{-n} (x) \, ,
\label{eq:S_re_T_W_Lag_m_1} \\
C  \left[ \sum_{n} \bar{\psi}_{n} (x) \psi_{n} (x) \right]  C^{-1}  
&= \sum_{n} \bar{\psi}_{n} \psi_{n} (x)
\, ,
\label{eq:C_W_Lag_m_1} \\
\Theta \left[ \sum_{n} \bar{\psi}_{n} (x) \psi_{n} (x) \right] \Theta^{-1}
&= \sum_{n} \bar{\psi}_{n} \psi_{n} (-x)
\, ,
\label{eq:CPT_W_Lag_m_1} 
\end{align}
which imply the invariance of the mass terms~\eqref{eq:LPsi_1}-\eqref{eq:LPsi_2}
\begin{align}
P \left[\overline{\Psi}(x) \Psi(x) \right] P^{-1} 
&=\overline{\Psi} \Psi (\mathscr{P}x) \, ,
\label{eq:P_W_Lag_m_2} \\
C  \left[\overline{\Psi}(x) \Psi(x) \right]  C^{-1}  
&=\overline{\Psi} \Psi (x)
\, ,
\label{eq:C_W_Lag_m_2} \\
C  P \left[\overline{\Psi}(x) \Psi(x) \right] P^{-1}  C^{-1}  
&=\overline{\Psi} \Psi (\mathscr{P}x)
\, ,
\label{eq:CP_W_Lag_m_2} \\
T \left[\overline{\Psi}(x) \Psi(x) \right] T^{-1} 
&=\overline{\Psi} \Psi (\mathscr{T}x) \, ,
\label{eq:T_W_Lag_m_2} \\
\hat{S}_{T} \left[\overline{\Psi}(x) \Psi(x) \right] \hat{S}_{T}^{-1} 
&= \overline{\Psi}(x) \Psi(x) \, ,
\label{eq:S_re_T_W_Lag_m_2} \\
\Theta \left[\overline{\Psi}(x) \Psi(x) \right] \Theta^{-1}
&=\overline{\Psi} \Psi (-x)
\, .
\label{eq:CPT_W_Lag_m_2} 
\end{align}
Combining these results above, we conclude that the full free Lagrangian density $\mathcal{L}_{0}(x)$~\eqref{eq:LPsi_1}-\eqref{eq:LPsi_2} remains invariant under $P$, $C$, $CP$, $T$, and $CPT$ transformations:
\begin{align}
P \mathcal{L}_{0}(x) P^{-1} 
&=\mathcal{L}_{0} (\mathscr{P}x) \, ,
\label{eq:P_W_Lag_0_2} \\
C  \mathcal{L}_{0}(x)  C^{-1}  
&=\mathcal{L}_{0} (x)
\, ,
\label{eq:C_W_Lag_0} \\
C  P \mathcal{L}_{0}(x) P^{-1}  C^{-1}  
&=\mathcal{L}_{0} (\mathscr{P}x)
\, ,
\label{eq:CP_W_Lag_0} \\
T \mathcal{L}_{0}(x) T^{-1} 
&=\mathcal{L}_{0} (\mathscr{T}x) \, ,
\label{eq:T_W_Lag_0_2} \\
\hat{S}_{T} \mathcal{L}_{0}(x) \hat{S}_{T}^{-1} 
&=\mathcal{L}_{0} (x) \, ,
\label{eq:S_re_T_W_Lag_0_2} \\
\Theta \mathcal{L}_{0}(x) \Theta^{-1}
&=\mathcal{L}_{0} (-x)
\, ,
\label{eq:CPT_W_Lag_0} 
\end{align}
where we have omitted a total derivative in the charge-conjugation~\eqref{eq:C_W_Lag_0} and $CPT$~\eqref{eq:CPT_W_Lag_0} transformations. 
Thus, the free action, which is the integral of the free Lagrangian density over spacetime, truly remains invariant under $P$, $C$, $CP$, $T$, and $CPT$ transformations separately. It is important to emphasize that this invariance holds independently of the concrete matrix forms for the charge-conjugation~\eqref{mat:C_state_W_ab_01} and $CPT$~\eqref{eq:cpt_matrix} transformations, indicating that the Wigner degeneracy does not affect the fundamental symmetry properties of the free theory. 
Although detailed analyses of typical configurations are provided in Sections~\ref{sec:C_W} and~\ref{sec:cpt_thm}, highlighting properties associated with the exchange of Wigner degeneracy in relation to these specific forms, these invariances still impose constraints on the interactions involving the Wigner degeneracy, which will be discussed in Section~\ref{sec:W_inter}.
Moreover, we can conclude that a free theory of Wigner doublet is always $\hat{S}_{T}$ invariant, 
Therefore, for a full theory including physical Wigner doublets, the internal $\hat{S}_{T}$ symmetry must be violated through some mechanism to be determined.


The Lagrangian formalism provides a natural framework to represent symmetries in the theory. Although the free Lagrangian density with the Wigner degeneracy~\eqref{eq:LPsi_1}-\eqref{eq:LPsi_2} is designed to be invariant under the continuous Lorentz group, the discrete $C$, $P$, $T$, and realize the correct Hamiltonian, rather than imposing by hand, it is accidentally invariant under the internal $U(2)$ transformation
\begin{align}
U(\beta) &= \exp \left( i \beta_{a} T^{a} \right) 
\, ,
\label{eq:U2_W_0}
\end{align}
where $T^{a}$, with $a=0,1,2,3$ are the $U(2)$ group generators. $\psi_{n}(x)$, $n=\pm\frac{1}{2}$ transform as
\begin{align}
\bigl[ T^{a}, \psi_{n} \bigr] = - \sum_{n'} \tau^{a}_{n n'} \psi_{n'} 
\, , \ \ 
\bigl[ T^{a}, \psi^{\dag}_{n} \bigr] = + \sum_{n'} \tau^{a*}_{n n'} \psi^{\dag}_{n'} 
\, , 
\label{eq:U2_W_1}
\end{align}
with $\tau^{a}=\frac{\sigma^{a}}{2}$, $a=0,1,2,3$, so that
\begin{align}
\bigl[ T^{a}, \mathcal{L}_{0}(x) \bigr] = 0 
\, , 
\label{eq:U2_Lag free_1}
\end{align}
due to the Hermitianity of $\tau^{a}$, leads to the invariance of the free Lagrangian density $\mathcal{L}_{0}(x)$~\eqref{eq:LPsi_1}-\eqref{eq:LPsi_2} under the internal $U(2)$ transformation manifestly
\begin{align}
U(\beta) \mathcal{L}_{0}(x) U^{-1}(\beta) = \mathcal{L}_{0}(x) 
\, .
\label{eq:U2_Lag free_2}
\end{align}
It also implies conserved currents and charges according to Noether's theorem. The Noether's current corresponding to each generator $T^{a}$ is given by
\begin{align}
J^{a \mu} = \sum_{n,n'} \bar{\psi}_{n} \gamma^{\mu} \tau^{a}_{n n'} \psi_{n'} = \overline{\Psi} \gamma^{\mu} \tau^{a} \Psi
\, , 
\ \ \text{with} \ \ 
\partial_{\mu} J^{a \mu} = 0 \, ,  
\label{eq:U2_Noe_free_1}
\end{align}
inducing the conserved charge
\begin{align}
Q^{a} = \int d^3x \ J^{a 0} (x) = \int d^3x \ \Psi^{\dag} (x) \tau^{a} \Psi (x)
\, .
\label{eq:U2_Noe_free_2}
\end{align}
Using the canonical anti-commutation relations~\eqref{eq:can_W_field_1}-\eqref{eq:can_W_field_2}, one can straightforward show that the charges $Q^{a}$, $a=0,1,2,3$~\eqref{eq:U2_Noe_free_2} satisfy the Lie algebra of $U(2)$ and transform the fields in the way of Eq.~\eqref{eq:U2_W_1} if we identify $Q^{a}=T^{a}$, with $a=0,1,2,3$. Thus, $Q^{a}$, $a=0,1,2,3$ can be treated as a concrete construction of the $U(2)$ generators in terms of fields.
Although the kinematic term is invariant under $U(2)$, the invariance of the mass term is coincidental. That is, the full $U(2)$ symmetry is preserved only when $\psi_{+\frac{1}{2}}(x)$ and $\psi_{-\frac{1}{2}}(x)$ have the degenerate mass. 
Note that the conserved charge associated with the $U(1)$ subgroup is given by
\begin{align}
Q^{0} = \int\frac{d^{3}p}{(2\pi)^{3}} \sum_{n, \sigma} 
\frac{1}{2} \left[ a_{n}^{\dag}(\p, \sigma) a_{n}(\p, \sigma) - a^{c \dag}_{n}(\p,\sigma) a^{c}_{n}(\p,\sigma) \right]
\, ,
\label{eq:U1_Noe_free_2}
\end{align}
which corresponds to the conservation of the Wigner number, where Wigner-degenerate particles contribute $+\frac{1}{2}$ to the Wigner number, while their antiparticles contribute $-\frac{1}{2}$.
Similarly, the conserved charge associated with the generator $T^{3}$ of the $SU(2)$ group is
\begin{align}
Q^{3} = \int\frac{d^{3}p}{(2\pi)^{3}} \sum_{n, \sigma} 
n\left[ a_{n}^{\dag}(\p, \sigma) a_{n}(\p, \sigma) - a^{c \dag}_{n}(\p,\sigma) a^{c}_{n}(\p,\sigma) \right]
\, ,
\label{eq:SU2_Noe_free_2}
\end{align}
which corresponds to the conservation of the Wigner charge. A Wigner-degenerate particle $|\p,\sigma,n;a\rangle$ contributes $n$ to the Wigner charge, while its corresponding antiparticle $|\p,\sigma,n;a^{c}\rangle$ contributes $-n$.
By combining the conservations of $Q^{0}$~\eqref{eq:U1_Noe_free_2} and $Q^{3}$~\eqref{eq:SU2_Noe_free_2}, we obtain individual Wigner charge conservations for two Wigner degeneracies $n=\pm\frac{1}{2}$ respectively, corresponding to the two redefined generators  
\begin{align}
T^{n} \equiv \frac{1}{2} \left[ T^{0} + (-)^{\frac{1}{2}-n} T^{3}  \right]
\, ,
\end{align}
which explicitly decouple the contributions of the Wigner doublet, providing a clearer perspective on their individual conservation properties.
If the Wigner doublet acquires two distinct masses, the global $U(2)$ symmetry is explicitly broken down to $U(1) \times U(1)$. It is intriguing that this explicit breaking of the internal $U(2)$ symmetry breaking necessarily implies the violation of either $CP$ or $CPT$ symmetry. 
This conclusion naturally extends to the case of an $n$-fold Wigner degeneracy, where similar symmetry-breaking patterns hold.
We notice that since $\hat{S}_{T} \in U(2)$, the $U(2)$ symmetry breaking inherently leads to the violation of $\hat{S}_{T}$ symmetry and thus the physical Wigner doublet can be identified.

\section{Interactions with the Wigner doublets}
\label{sec:W_inter}

According to the well-known $CPT$ theorem in CQFT, one can always choose proper intrinsic discrete symmetry phases so that the $CPT$ transformation $\Theta$ leaves the interacting Hamiltonian density $\mathcal{H}_{I} (x)$ invariant~\cite{Weinberg:1995mt}:
\begin{equation}
    \Theta\mathcal{H}_{I}(x)\Theta^{-1}=\mathcal{H}_{I}(-x)\,.
    \label{eq:cpt_H_I}
\end{equation}
As a result, the interaction potential $V \equiv \int d^{3}x \ \mathcal{H}_{I}(0, \mathbf{x})$ commutes with $\Theta$, ensuring the $CPT$ invariance.
However, in the presence of Wigner degeneracy, the mixing induced by $CPT$ transformations introduces additional complications. This necessitates a more careful and detailed analysis to fully understand how $CPT$ is invariant or breaking in such a framework. The results in Section~\ref{sec:cpt_thm} provide fundamental building blocks for this derivation.
In this section, we study nontrivial interacting models within the framework of two-fold Wigner QFT. Our approach involves a critical examination of the model construction, the imposition of physical constraints, and a systematic narrowing of viable possibilities.
While the criterion of Eq.~\eqref{eq:cpt_H_I} still indicates the $CPT$ invariance in Wigner QFT, the interaction terms involving the Wigner doublets must be carefully formulated using the bilinear structures analyzed in the previous sections. 
However,  these terms can also be carefully designed to explicitly violate Eq.~\eqref{eq:cpt_H_I}, leading to an explicit breaking of $CPT$ invariance in the presence of Wigner degeneracy. 
Moreover, as highlighted at the end of Section~\ref{sec:PT_1}, a physically meaningful two-fold Wigner model induced by time-reversal symmetry can only be realized if there exists no internal symmetry operator $\hat{S}_{T}$~\eqref{eq:S_re_T_0}, which will provide guidance in our model construction.
To illustrate our points concretely, we focus on two typical interacting models in the subsequent discussions.
\\

\subsection{Yukawa interaction} 
\label{sec:W_inter_Yuk}

Let $X (x)$, $\Psi (x)$ be two Wigner doublet fields, defined as 
\begin{align}
X(x) 
\equiv\left[\begin{matrix}
    \chi_{+\frac{1}{2}}(x) \\
    \chi_{-\frac{1}{2}}(x)
\end{matrix}\right] 
\, ,
\quad
\Psi(x)
\equiv\left[\begin{matrix}
    \psi_{+\frac{1}{2}}(x) \\
    \psi_{-\frac{1}{2}}(x)
\end{matrix}\right] 
\, .
\end{align}
The generic Yukawa interaction between two Wigner doublets and a complex scalar field $\phi(x)$ can be written in the doublet form as
\begin{equation}
\mathcal{H}_{Yuk}(x;Y)= \overline{X} (x) Y \Psi (x) \phi (x) + \text{h.c.} \, ,
\label{eq:Yuk_W_2_1}
\end{equation}
where $Y$ is the Yukawa coupling matrix.
To maintain generality, the Yukawa coupling matrix is taken to be chiral and complex
\begin{equation}
Y_{mn} \equiv
y^{L}_{mn} P_{L} + y^{R}_{mn} P_{R}
\, .
\label{mat:Yuk_W_2_1}
\end{equation}
In the context of Wigner QFT, the $CPT$ transformation for the Yukawa interaction can be derived by applying the results from bilinears~\eqref{eq:CPT_W_bi_1} with $\Gamma=Y_{mn}$, $r=4$:
\begin{align}
    \Theta\mathcal{H}_{Yuk}(x;Y)\Theta^{-1}&=\mathcal{H}_{Yuk}(-x;Y')\,,\label{eq:cpt_H_phi}
\end{align}
where the transformed Yukawa coupling matrix $Y'$ is given by
\begin{equation}
Y^{\prime \text{T}} \equiv \Xi Y^{\text{T}} \widetilde{\Xi}^{\dag} \, ,
\label{eq:Yuk_CPT_1}
\end{equation}
and we conventionally set $\Theta\phi(x)\Theta^{-1}=\phi^{\dag}(-x)$ as in CQFT. 
It implies that, in general, $Y'\neq Y$. Thus, the Yukawa interaction is not necessarily invariant after $CPT$, and this violation arises due to the complex mixing of the Wigner doublets. 
Besides, the different $CPT$ matrices associated with the two Wigner doublets can introduce additional modifications to the transformed Yukawa couplings.
In CQFT, where the Wigner degeneracy is absent, $CPT$ invariance always holds without any such complications.
However, in the Wigner doublet framework, the interplay of these degeneracies results in an intrinsic distinction between the Lagrangian before and after the $CPT$ transformation. This results in a scenario where $CPT$ symmetry can be violated, revealing the unique characteristics of the Wigner doublet framework. Such mixing not only increases the theoretical complexity but also introduces new and rich phenomenological implications, deepening our understanding of symmetry and interactions in QFT.

Furthermore, the time-reversal on transformation for the Yukawa interaction can be derived by applying the results for bilinears~\eqref{eq:T_W_bi_1} with $\Gamma=Y_{mn}$:
\begin{align}
T \mathcal{H}_{Yuk}(x;Y) T^{-1}
= \overline{X}(\mathscr{T}x) 
\left[ \tilde{D}(\mathscr{T}) Y^{*} D(\mathscr{T}) \right]
\Psi(\mathscr{T}x)
\phi (\mathscr{T}x) 
+ \text{h.c.}
\label{eq:Yuk_T_1}
\, ,
\end{align}
and its associated $\hat{S}_{T}$ transformation~\eqref{eq:S_re_T_field_2} for the Yukawa interaction is given as
\begin{align}
\hat{S}_{T} \mathcal{H}_{Yuk}(x;Y) \hat{S}_{T}^{-1}
= \overline{X}(x) 
\left[ \tilde{D}(\mathscr{T}) Y D(\mathscr{T}) \right] \Psi(x)
\phi (x)
+ \text{h.c.}
\, ,
\label{eq:Yuk_ST_1}
\end{align}
where we have imposed the result $\hat{S}_{T} \phi (x) \hat{S}_{T}^{-1} = \phi (x)$ as in CQFT.
As a concrete example, we construct a Yukawa interaction that violates $CPT$ symmetry but preserves the time-reversal symmetry without an internal $\hat{S}_{T}$~\eqref{eq:S_re_T_0}. 
Combining Eqs.~\eqref{eq:Yuk_CPT_1}, \eqref{eq:Yuk_T_1} and \eqref{eq:Yuk_ST_1}, we find that the Yukawa couplings must satisfy the conditions:
\begin{align}
Y^{\text{T}} \neq \Xi Y^{\text{T}} \widetilde{\Xi}^{\dag} 
\, , \quad
Y = \tilde{D}(\mathscr{T}) Y^{*} D(\mathscr{T}) 
\, , \quad
Y \neq \tilde{D}(\mathscr{T}) Y D(\mathscr{T}) 
\, .
\label{eq:Yuk_con_T_1}
\end{align}
For simplicity, we assume universal transformation matrices for all the Wigner doublets:
\begin{align}
\widetilde{\Xi} = \Xi 
= \left[\begin{matrix}
        0 & e^{i\frac{\varphi}{2}} \\
        e^{-i\frac{\varphi}{2}} & 0
    \end{matrix}\right]
\, , \quad
\tilde{D}(\mathscr{T}) = D(\mathscr{T}) 
= \left[\begin{matrix}
        0 & e^{i\frac{\phi}{2}} \\
        e^{-i\frac{\phi}{2}} & 0
\end{matrix}\right]
\, ,
\end{align}
along with a diagonal Yukawa coupling matrix
\begin{align}
y^{L,R} = 
\left[\begin{matrix}
        y_{+\frac{1}{2}, +\frac{1}{2}}^{L,R} & 0 \\
        0 & y_{-\frac{1}{2}, -\frac{1}{2}}^{L,R}
\end{matrix}\right] \, .
\end{align}
The condition in Eq.~\eqref{eq:Yuk_con_T_1} then explicitly constrains the Yukawa couplings as
\begin{align}
y_{+\frac{1}{2}, +\frac{1}{2}}^{L,R} = y_{-\frac{1}{2}, -\frac{1}{2}}^{L,R*} \notin \mathbb{R} \, .
\end{align}
Thus, this Yukawa coupling structure explicitly breaks $CPT$ symmetry but preserves the time-reversal invariance. 
More significantly, it eliminates the internal symmetry responsible for diagonalizing the time-reversal operator
so that the Wigner doublet degenerated by the time-reversal symmetry is truly physical. 
We need to emphasize that despite the explicit breaking of $\hat{S}_{T}$, a residue global symmetry $U(1) \times U(1)$ remains valid, so that the two charges $Q^{0}$~\eqref{eq:U1_Noe_free_2} and $Q^{3}$~\eqref{eq:SU2_Noe_free_2} are still conserved.
In addition, this unique Yukawa coupling structure within the Wigner doublet framework resolves the ambiguity problem in space-time reflection operators previously discussed in Ref.~\cite{Lee:1966ik}.

However, the $CPT$ invariance of the Yukawa interactions~\eqref{eq:Yuk_W_2_1} can be occasionally restored if 
\begin{align}
\bigl[Y,\Xi^{\text{T}} \bigr]&=0 \, , 
\ \  
\widetilde{\Xi}=\Xi \, .
\label{eq:CPT_yuk_con_1}
\end{align}
Since $\Xi$ is an arbitrary constant $U(2)$ matrix as factorized explicitly in Eq.~\eqref{eq:cpt_matrix}, one trivial solution which preserves the $CPT$ symmetry is $\widetilde{\Xi}=\Xi=\mathds{1}$, which implies no Wigner mixing under the $CPT$ transformation.
Alternatively, the Yukawa coupling matrix $Y$~\eqref{mat:Yuk_W_2_1} can be reduced to a non-chiral form $Y_{nm}=y\delta_{nm}$, with $y\in\mathbb{C}$, which also satisfies the commutation condition~\eqref{eq:CPT_yuk_con_1}.
Beyond these trivial cases, nontrivial solutions to Eq.~\eqref{eq:CPT_yuk_con_1} might be $Y= y \Xi^{\text{T}}$ or $y \Xi^{*}$, $y\in\mathbb{C}$.
These solutions, derived from the unitarity of $\Xi$~\eqref{eq:cpt_matrix}, generally lead to Yukawa interactions that induce an exchange between the Wigner doublets.

To make the first cut as a DM candidate, the Wigner doublets must be electrically neutral. 
The absence of electric charge would limit their interactions with SM leptons $\ell_{SM}$ mediated by the Higgs boson. Specifically, an interaction of the form $\bar{\psi}_{n}\ell_{SM}\phi$ is forbidden, as it would induce the Higgs decay process $H\rightarrow\bar{\psi}_{n}\ell_{SM}$ which violates charge conservation. 
A promising strategy for detecting and constraining the Wigner doublet is through proton-proton collision at the Large Hadron Collider (LHC).
Assuming that the Higgs boson couples to the Wigner doublet via Eq.~(\ref{eq:Yuk_W_2_1}), the Higgs boson can then decay to these fermions, which constitute an invisible channel. A key process to probe this scenario is proton-proton collision producing a Higgs boson and a $Z$ boson $pp\rightarrow HZ$ with $H\rightarrow\bar{X}\Psi$, $H\rightarrow\bar{\Psi}X$~\cite{CMS:2016dhk,2018318}. The Yukawa interactions provide a potential avenue for identifying the Wigner doublets that interact with the SM leptons mediated by the Higgs boson. 
If the two Wigner-degenerate doublets $X$ and $\Psi$ carry different conserved charges, then interaction terms like $\overline{X}Y\Psi H$ would be forbidden by charge conservation. As a result, the Yukawa interactions are constrained to $\overline{\Psi}Y\Psi H$, $\overline{X}YXH$. The Higgs decay channels are limited to $H\rightarrow\overline{\Psi}\Psi$, $H\rightarrow\overline{X}X$. 
For instance, these fermions could be assigned $U(2)$ (or residue) charges as discussed in the previous section. This naturally raises the question of whether a $U(2)$ gauge interaction can be incorporated into the framework, which we now proceed to explore. 






\subsection{Gauge interactions} 
\label{sec:W_inter_G}

As demonstrated in Section~\ref{sec:symmetry_W_free}, the free Lagrangian density of the Wigner doublet~\eqref{eq:LPsi_1}-\eqref{eq:LPsi_2} accidentally exhibits a non-chiral global $U(2)$ symmetry.  
In order to gauge the $U(2)$ symmetry, which can be decomposed as $U(2)=U(1) \times SU(2)/ \mathbb{Z}_{2}$, one need to introduce a gauge field $V_{\mu}(x)$ for the $U(1)$ sector and three other gauge fields $G^{a}_{\mu}(x)$, $a=1,2,3$ for the $SU(2)$ sector.
These gauge fields couple to the Wigner doublet fields through minimal coupling, giving rise to the Wigner doublet-gauge interaction terms
\begin{align}
\mathcal{H}_{gauge} = \mathcal{H}_{A} + \mathcal{H}_{YM} \, ,
\label{eq:G_U2_W_1}
\end{align}
where
\begin{align}
\mathcal{H}_{A}
&= g J_{A}^{\mu} V_{\mu} \, ,
\label{eq:G_U1_W_1} \\
\mathcal{H}_{YM}
&= g' \sum_{a} J_{YM}^{a \mu} G^{a}_{\mu} 
\, ,
\label{eq:G_SU2_W_1}
\end{align}
with the gauge coupling constants $g, g'\in\mathbb{R}$ and the corresponding currents
\begin{align}
J_{A}^{\mu}
&= \overline{\Psi} \gamma^{\mu} \tau^{0} \Psi 
= \frac{1}{2} \sum_{n}\bar{\psi}_{n} \gamma^{\mu} \psi_{n} \, ,
\label{eq:G_U1_W_J_1} \\
J_{YM}^{a \mu}
&= \overline{\Psi} \gamma^{\mu} \tau^{a} \Psi
= \sum_{n,n'}\bar{\psi}_{n} \gamma^{\mu}\tau^{a}_{nn'} \psi_{n'} 
\, ,
\label{eq:G_SU2_W_J_1}
\end{align}
where $\tau^{0}=\frac{\sigma^{0}}{2}$ is the generator of $U(1)$, and $\tau^{a}=\frac{\sigma^{a}}{2}$, $a=1,2,3$ are the three generators of $SU(2)$.
It is obvious that the two gauge fields 
$V_{\mu}$ and $G^{3}_{\mu}$ couple to the Wigner neutral currents, preserving the Wigner degeneracy. In contrast, the gauge interactions mediated by $G^{1}_{\mu}$ and $G^{2}_{\mu}$ induce transitions between different Wigner-degenerate states. To gain further insight, we can reformulate the Wigner doublet-gauge interaction terms by rotating the gauge field basis
\begin{align}
\left[\begin{matrix}
        \mathcal{A}_{\mu} \\
        \mathcal{B}_{\mu} 
\end{matrix}\right]
&\equiv
\left[\begin{matrix}
        \cos \theta_{D} & \sin \theta_{D} \\
        -\sin \theta_{D} & \cos \theta_{D}
\end{matrix}\right]
\left[\begin{matrix}
        V_{\mu} \\
        G^{3}_{\mu}
\end{matrix}\right]
\, ,
\label{eq:G_U2_W_2_2n} \\
\mathcal{G}^{+}_{\mu} &\equiv \frac{G^{1}_{\mu} - i G^{2}_{\mu}}{\sqrt{2}}
\, , \ \ 
\mathcal{G}^{-}_{\mu} \equiv \frac{G^{1}_{\mu} + i G^{2}_{\mu}}{\sqrt{2}} \, ,
\label{eq:G_U2_W_2_2W}
\end{align}
with
\begin{align}
\sin \theta_{D} = \dfrac{g}{g_{D}} \, , \ \ 
\cos \theta_{D} = \dfrac{g'}{g_{D}} \, , \ \ 
g_{D} \equiv \sqrt{g^{2} + g^{\prime 2}} \, .
\label{eq:G_U2_W_2_g}
\end{align}
The Wigner doublet-gauge interaction $\mathcal{H}_{gauge}$~\eqref{eq:G_U2_W_1}-\eqref{eq:G_SU2_W_1} can then be decomposed as
\begin{align}
\mathcal{H}_{gauge} = \mathcal{H}_{\mathcal{A}} + \mathcal{H}_{\mathcal{B}} + \mathcal{H}_{\mathcal{G}} \, ,
\label{eq:G_U2_W_2}
\end{align}
where
\begin{alignat}{2}
\mathcal{H}_{\mathcal{A}} 
&= \dfrac{gg'}{g_{D}} J_{\mathcal{A}}^{\mu} \mathcal{A}_{\mu}
\, , \quad &
J_{\mathcal{A}}^{\mu} 
&= \bar{\psi}_{+\frac{1}{2}}\gamma^{\mu} \psi_{+\frac{1}{2}}
\, ,
\label{eq:G_U2_W_2_A} \\
\mathcal{H}_{\mathcal{B}} 
&= \dfrac{g_{D}}{2} J_{\mathcal{B}}^{\mu} \mathcal{B}_{\mu}
\, , \quad &
J_{\mathcal{B}}^{\mu}
&= \left( 1- 2\sin^{2} \theta_{D} \right) J_{\mathcal{A}}^{\mu} 
- \bar{\psi}_{-\frac{1}{2}}\gamma^{\mu} \psi_{-\frac{1}{2}}
\, ,
\label{eq:G_U2_W_2_B} \\
\mathcal{H}_{\mathcal{G}} 
&= \dfrac{g'}{\sqrt{2}} 
\left(J_{\mathcal{G}}^{\mu \dagger} \mathcal{G}^{+}_{\mu}
+ J_{\mathcal{G}}^{\mu} \mathcal{G}^{-}_{\mu}
\right)
\, , \quad & 
J_{\mathcal{G}}^{\mu} 
&= \bar{\psi}_{-\frac{1}{2}}\gamma^{\mu} \psi_{+\frac{1}{2}}
\, .
\label{eq:G_U2_W_2_W}
\end{alignat}
The gauge fields $\mathcal{G}^{\pm}$~\eqref{eq:G_U2_W_2_2W} couple to the Wigner charged currents $J_{\mathcal{G}}^{\mu}$~\eqref{eq:G_U2_W_2_W}, responsible for raising and lowering the Wigner indices. 
Meanwhile, gauge fields $\mathcal{A}_{\mu}$ and $\mathcal{B}_{\mu}$~\eqref{eq:G_U2_W_2_2n} couple to the Wigner neutral currents $J_{\mathcal{A}}^{\mu}$~\eqref{eq:G_U2_W_2_A} and $J_{\mathcal{B}}^{\mu}$~\eqref{eq:G_U2_W_2_B} respectively, preserving the Wigner index. 
Notice that $\mathcal{A}_{\mu}$ interacts exclusively with the $n=+\frac{1}{2}$ Wigner current.
Moreover, if $|g|=|g'|$ is satisfied in Eq.~\eqref{eq:G_U2_W_2_g}, then $\mathcal{B}_{\mu}$ will only couple to the $n=-\frac{1}{2}$ Wigner current, ensuring that the Wigner neutral currents for $n=\pm\frac{1}{2}$ decouple from each other.

Next, let us examine the $CPT$ invariance of the gauge sector within the Wigner doublet framework.
We first consider the $CPT$ invariance in the $U(1)$ gauge sector~\eqref{eq:G_U1_W_1} alone. A straightforward calculation with Eq.~\eqref{eq:CPT_W_Lag_kin_1}, gives the $CPT$ transformation
\begin{equation}
    \Theta\mathcal{H}_{A}(x)\Theta^{-1}=\mathcal{H}_{A}(-x)\,,
    \label{eq:CPT_U2_W_U1_A}
\end{equation}
where we have imposed $\Theta V_{\mu}(x)\Theta^{-1}=-V_{\mu}(-x)$ as in CQFT. 
This result directly implies that the $U(1)$ gauge potential
\begin{equation}
V_{A}=\int d^{3}x \  \mathcal{H}_{A}(0,\x) \, ,
\end{equation}
commutes with $CPT$ [i.e., $\Theta V_{A} \Theta^{-1} = V_{A}$] and thus preserves $CPT$ symmetry.
However, taking into account the non-Abelian $SU(2)$ gauge sector~\eqref{eq:G_SU2_W_1} will disrupt the $CPT$ symmetry. This violation becomes evident in the basis presented in Eqs.~\eqref{eq:G_U2_W_2_A}-\eqref{eq:G_U2_W_2_W}. 
In particular, $\mathcal{H}_{\mathcal{A}}$~\eqref{eq:G_U2_W_2_A}, which only couples to the $n=+\frac{1}{2}$ Wigner current, explicitly breaks the general $CPT$ symmetry, since the $CPT$ transformation generally mixes the Wigner degeneracies.
Thus, the $CPT$ symmetry breaking can only be avoided if additional constraints are imposed to enforce Wigner symmetry between the two Wigner neutral gauge fields $\mathcal{A}_{\mu}(x)$ and $\mathcal{B}_{\mu}(x)$. 
To illustrate how such a process can be implemented, we construct a simplified model by assuming uniform gauge couplings, given by
\begin{align}
g = g' \, , \ \ 
\sin \theta_{D} = \cos \theta_{D} = \frac{\sqrt{2}}{2} 
\, ,
\label{eq:G_U2_W_2_gg}
\end{align}
so that 
\begin{align}
\left[\begin{matrix}
        \mathcal{A}_{\mu} \\
        \mathcal{B}_{\mu} 
\end{matrix}\right]
= \frac{\sqrt{2}}{2}
\left[\begin{matrix}
        1 & 1 \\
        -1 & 1
\end{matrix}\right]
\left[\begin{matrix}
        V_{\mu} \\
        G^{3}_{\mu}
\end{matrix}\right]
\, .
\label{eq:G_U2_W_2_2n_2} 
\end{align}
The Wigner doublet-gauge interactions in Eqs.~\eqref{eq:G_U2_W_2}-\eqref{eq:G_U2_W_2_W} can be further simplified to
\begin{alignat}{2}
\mathcal{H}_{\mathcal{A}} 
&= \dfrac{g}{\sqrt{2}} J_{\mathcal{A}}^{\mu} \mathcal{A}_{\mu}
\, , \quad & 
J_{\mathcal{A}}^{\mu} 
&= \bar{\psi}_{+\frac{1}{2}}\gamma^{\mu} \psi_{+\frac{1}{2}}
\, ,
\label{eq:G_U2_W_2_A_gg} \\
\mathcal{H}_{\mathcal{B}} 
&= \dfrac{g}{\sqrt{2}} J_{\mathcal{B}}^{\mu} \mathcal{B}_{\mu}
\, , \quad &
J_{\mathcal{B}}^{\mu}
&= - \bar{\psi}_{-\frac{1}{2}}\gamma^{\mu} \psi_{-\frac{1}{2}}
\, ,
\label{eq:G_U2_W_2_B_gg} \\
\mathcal{H}_{\mathcal{G}} 
&= \dfrac{g}{\sqrt{2}} 
\left(J_{\mathcal{G}}^{\mu \dagger} \mathcal{G}^{+}_{\mu}
+ J_{\mathcal{G}}^{\mu} \mathcal{G}^{-}_{\mu}
\right)
\, , \quad& 
J_{\mathcal{G}}^{\mu} 
&= \bar{\psi}_{-\frac{1}{2}}\gamma^{\mu} \psi_{+\frac{1}{2}}
\, .
\label{eq:G_U2_W_2_W_gg}
\end{alignat}
To achieve a nontrivial result, we consider a $CPT$ transformation that exchanges the Wigner degeneracies, which is characterized by an anti-diagonal matrix $\Xi$ as defined in Eq.~\eqref{eq:cpt_matrix2}.
By selecting appropriate phases $\varphi=0$, the $CPT$ transformation matrix $\Xi$ takes the explicit form
\begin{align}
\Xi =
\left[\begin{matrix}
0 \ \  & \ \ 1  \\
1 \ \  & \ \ 0  
\end{matrix}\right]
\, ,
\label{eq:G_U2_W_gg_Xi}
\end{align}
which directly exchanges the Wigner degeneracies without introducing additional phase factors. 
Applying the general $CPT$ transformations on bilinears~\eqref{CPT_W_bi_bd_1} with $\Gamma=\gamma^{\mu}$, $r=1$,
we obtain the explicit transformation of the Wigner neutral currents
$J_{\mathcal{A}}^{\mu}(x)$~\eqref{eq:G_U2_W_2_A_gg} and $J_{\mathcal{B}}^{\mu}(x)$~\eqref{eq:G_U2_W_2_B_gg}
under the specific $CPT$ transformation~\eqref{eq:G_U2_W_gg_Xi}:
\begin{align}
\Theta J_{\mathcal{A}}^{\mu}(x) \Theta^{-1}
&= J_{\mathcal{B}}^{\mu}(-x)
\, , \ \  
\Theta J_{\mathcal{B}}^{\mu}(x) \Theta^{-1}
= J_{\mathcal{A}}^{\mu}(-x)
\, , 
\label{eq:CPT_U2_N_gg_J_1} 
\end{align}
which confirms that $CPT$~\eqref{eq:G_U2_W_gg_Xi} exchanges $J_{\mathcal{A}}^{\mu}$ and $J_{\mathcal{B}}^{\mu}$ as expected.
Thus, the $CPT$ invariance of the Wigner neutral sector, described by the interaction Hamiltonian $\mathcal{H}_{\mathcal{A}} + \mathcal{H}_{\mathcal{B}}$~\eqref{eq:G_U2_W_2_A_gg}-\eqref{eq:G_U2_W_2_B_gg}
\begin{equation}
    \Theta \left[ \mathcal{H}_{\mathcal{A}}(x) + \mathcal{H}_{\mathcal{B}}(x) \right]\Theta^{-1}
    = \mathcal{H}_{\mathcal{A}}(-x) + \mathcal{H}_{\mathcal{B}}(-x) \, ,
    \label{eq:CPT_U2_W_U2_N}
\end{equation}
requires the dramatic Wigner symmetry between the Wigner neutral gauge fields $\mathcal{A}_{\mu}(x)$ and $\mathcal{B}_{\mu}(x)$:
\begin{align}
\Theta
\left[\begin{matrix}
        \mathcal{A}_{\mu}(x) \\
        \mathcal{B}_{\mu}(x)
\end{matrix}\right]
\Theta^{-1}
=\left[\begin{matrix}
        \mathcal{B}_{\mu}(-x) \\
        \mathcal{A}_{\mu}(-x)
\end{matrix}\right]
\label{eq:CPT_U2_N_gg_field_1}
\, .
\end{align}
Furthermore, using the basis transformation~\eqref{eq:G_U2_W_2_2n_2},
this Wigner symmetry structure leads to the $CPT$ transformation properties of the gauge fields $V_{\mu}(x)$ and $G^{3}_{\mu}(x)$:  
\begin{align}
\Theta
\left[\begin{matrix}
        V_{\mu}(x) \\
        G^{3}_{\mu}(x)
\end{matrix}\right]
\Theta^{-1}
=\left[\begin{matrix}
        -V_{\mu}(-x) \\
        G^{3}_{\mu}(-x)
\end{matrix}\right]
\label{eq:CPT_U2_N_gg_field_2}
\, ,
\end{align}
which implies that $V_{\mu}(x)$ transforms in the same way as in the pure $U(1)$ gauge case given in Eq.~\eqref{eq:CPT_U2_W_U1_A}. 
For the Wigner charged sector $\mathcal{H}_{\mathcal{G}}$~\eqref{eq:G_U2_W_2_W_gg},
we obtain the explicit $CPT$ transformation of the Wigner charged currents $J_{\mathcal{G}}^{\mu}(x)$~\eqref{eq:G_U2_W_2_W_gg} under the specific $CPT$ matrix~\eqref{eq:G_U2_W_gg_Xi}:
\begin{align}
\Theta J_{\mathcal{G}}^{\mu}(x) \Theta^{-1}
&= -J_{\mathcal{G}}^{\mu}(-x)
\, , \ \ 
\Theta J_{\mathcal{G}}^{\mu \dag}(x) \Theta^{-1}
= -J_{\mathcal{G}}^{\mu \dag}(-x)
\, .
\label{eq:CPT_U2_W_gg_J_1}
\end{align}
As a result, the $CPT$ invariance of the Wigner charged sector
\begin{equation}
    \Theta\mathcal{H}_{\mathcal{G}}(x)\Theta^{-1}
    = \mathcal{H}_{\mathcal{G}}(-x)\, ,
    \label{eq:CPT_U2_W_U2_W}
\end{equation}
requires the Wigner charged gauge fields $\mathcal{G}^{\pm}_{\mu}(x)$ transformed under $CPT$ as
\begin{align}
\Theta \mathcal{G}^{\pm}_{\mu}(x) \Theta^{-1}
=-\mathcal{G}^{\pm}_{\mu}(-x)
\, .
\label{eq:CPT_U2_W_gg_field_1}
\end{align}
Then, reversing the basis transformation in Eq.~\eqref{eq:G_U2_W_2_2W}, we obtain the corresponding $CPT$ transformations for the gauge fields $G^{1}_{\mu}(x)$ and $G^{2}_{\mu}(x)$:
\begin{align}
\Theta
\left[\begin{matrix}
        G^{1}_{\mu}(x) \\
        G^{2}_{\mu}(x)
\end{matrix}\right]
\Theta^{-1}
&=
\left[\begin{matrix}
        -G^{1}_{\mu}(-x) \\
        G^{2}_{\mu}(-x)
\end{matrix}\right]
\label{eq:CPT_U2_W_gg_field_2}
\, .
\end{align}
At this stage, we have constructed a specific $CPT$ invariant model 
\begin{align}
\Theta \mathcal{H}_{gauge}(x) \Theta^{-1}
= \mathcal{H}_{gauge}(-x) \, ,
\label{eq:CPT_U2_W_U2_G}
\end{align}
by preserving the $CPT$ invariant in the Wigner neutral sector $\mathcal{H}_{\mathcal{A}} + \mathcal{H}_{\mathcal{B}}$~\eqref{eq:CPT_U2_W_U2_N} and the Wigner charged sector $\mathcal{H}_{\mathcal{G}}$~\eqref{eq:CPT_U2_W_U2_W} separately.
While $CPT$ is generally not invariant in the gauge sector $\mathcal{H}_{gauge}$~\eqref{eq:G_U2_W_1} due to the presence of Wigner degeneracy, it can be recovered if the gauge fields satisfy a specific Wigner symmetry under $CPT$ transformations.
The Wigner symmetry of the gauge fields under a general $CPT$ can be derived by inserting Eq.~\eqref{eq:CPT_W_bi_1} with $\Gamma=\gamma^{\mu}$, $r=1$ to Eq.~\eqref{eq:CPT_U2_W_U2_G}
\begin{align}
V_{\mu}(x) &\rightarrow V'_{\mu}(-x) = -V_{\mu}(-x) 
\, , \\
\sum_{a} G^{a}_{\mu}(x) \tau^{a}
&\rightarrow \sum_{a} G^{\prime a}_{\mu}(-x) \tau^{a}
= -\sum_{a} G^{a}_{\mu}(-x) \Xi^{\text{T}} \tau^{a} \Xi^{*}
\, ,
\end{align}
which reduce to our previous results in Eqs.~\eqref{eq:CPT_U2_N_gg_field_2} and \eqref{eq:CPT_U2_W_gg_field_2} under the specific $\Xi$~\eqref{eq:G_U2_W_gg_Xi}.
Furthermore, the time-reversal on transformation for the gauge sector $\mathcal{H}_{gauge}$~\eqref{eq:G_U2_W_1} can be derived by applying the results for bilinears~\eqref{eq:T_W_bi_1} with $\Gamma=\gamma^{\mu}$:
\begin{align}
T \mathcal{H}_{A}(x) T^{-1}
&= g \overline{\Psi}(\mathscr{T}x) 
\eta_{\mu \nu} \gamma^{\nu}
T \left[V_{\mu}(x) \tau^{0} \right] T^{-1} \Psi(\mathscr{T}x)
\, , \\
T \mathcal{H}_{YM}(x) T^{-1}
&= g' \overline{\Psi}(\mathscr{T}x) 
\eta_{\mu \nu} \gamma^{\nu}
D(\mathscr{T})
T \left[ \sum_{a} G^{a}_{\mu}(x) \tau^{a} \right] T^{-1}
D(\mathscr{T})
\Psi(\mathscr{T}x)
\, .
\end{align}
The time-reversal invariance of the gauge sector $\mathcal{H}_{gauge}$~\eqref{eq:G_U2_W_1}
\begin{align}
T \mathcal{H}_{gauge}(x) T^{-1}
= \mathcal{H}_{gauge}(\mathscr{T}x) \, ,
\label{eq:T_U2_W_U2_G}
\end{align}
requires that
\begin{align}
V_{\mu}(x) &\rightarrow V'_{\mu}(\mathscr{T}x) = V^{\mu}(\mathscr{T}x) 
\, , \\
\sum_{a} G^{a}_{\mu}(x) \tau^{a}
&\rightarrow \sum_{a} G^{\prime a}_{\mu}(\mathscr{T}x) \tau^{a}
= \sum_{a} G^{a \mu}(\mathscr{T}x) D(\mathscr{T}) \tau^{a} D(\mathscr{T})
\, .
\end{align}
The solutions of $G^{\prime a}_{\mu}$, $a=1,2,3$ are given by inserting the explicit form of $D(\mathscr{T})$~\eqref{mat:T_state_W_ab_2}:
\begin{equation}
\left\{
\begin{array}{r c l}
G^{\prime 1}_{\mu} (\mathscr{T}x) &=& \cos \phi \, G^{1 \mu} (\mathscr{T}x) 
- \sin \phi \, G^{2 \mu} (\mathscr{T}x) \, , 
\\ \vspace{-0.2cm} \\
G^{\prime 2}_{\mu} (\mathscr{T}x) &=& - \sin \phi \, G^{1 \mu} (\mathscr{T}x) - \cos \phi \, G^{\prime 2 \mu} (\mathscr{T}x) \, .
\end{array}
\right.
\end{equation}
Thus, similar to the $CPT$ invariance, the time-reversal invariance of the gauge sector also requires a Wigner symmetry of the gauge fields, which also arises from the mixing of Wigner degeneracies.
As we proposed in Section~\ref{sec:PT_1}, if the Wigner doublet is physically meaningful, the internal symmetry $\hat{S}_{T}$~\eqref{eq:S_re_T_0} must be absent. However, since $\hat{S}_{T}$ is an element of $U(2)$ group and the gauge sector is invariant under the global $U(2)$ group, the Lagrangian must be invariant under $\hat{S}_{T}$.
Nevertheless, a spontaneous $U(2)$ symmetry breaking may occur if the vacuum expectation values of the Wigner doublet are not equal
\begin{align}
\langle \Omega | \bar{\psi}_{+\frac{1}{2}} \psi_{+\frac{1}{2}} | \Omega \rangle 
= V_{+\frac{1}{2}}^{3}
\neq
V_{-\frac{1}{2}}^{3} = 
\langle \Omega | \bar{\psi}_{-\frac{1}{2}} \psi_{-\frac{1}{2}} | \Omega \rangle
\, ,
\end{align}
where $| \Omega \rangle$ is the true (stable) vacuum, so that the $U(2)$ global symmetry would be broken down to $U(1) \times U(1)$.
The unbroken symmetry corresponds to the individual Wigner charge subgroup, which in turn implies the decoherence of the Wigner doublet. 
This spontaneous symmetry breaking induces a phase transition~\footnote{It may be just a crossover in a strict sense.}, after which the internal symmetry $\hat{S}_{T}$, which mixes different Wigner degeneracies, is no longer preserved, i.e., $\hat{S}_{T} | \Omega \rangle \neq | \Omega \rangle$.
As a result, the two components of the Wigner doublet no longer mix coherently and become physically distinguishable states. Such a symmetry structure ensures that interactions respect the separation of Wigner-degenerate states, preventing superpositions that would otherwise obscure their distinct physical roles.
This scenario may have some phenomenological implications that can be observed in future experiments. In big bang cosmology, the spontaneously broken $U(2)$ symmetry may be restored at sufficiently high temperature~\cite{Kirzhnits:1972iw,Dolan:1973qd,Weinberg:1974hy,Linde:1978px}, potentially triggering a phase transition in the early universe.
If this transition is of the first order, it implies that just below a critical temperature, the universe would evolve from a metastable quasi-equilibrium state to a stable equilibrium state via bubble nucleation.
Such a process can give rise to a variety of observable signatures.
For instance, if the first-order phase transition occurs during the cosmic inflation, it may lead to the formation of topological defects~\cite{Vilenkin:1981zs,Vilenkin:1982ks,Kibble:1982dd,Kibble:1984hp,Vilenkin:1984ib,Hindmarsh:1994re}, which can imprint characteristic anisotropies on the cosmic microwave background (CMB). If it occurs near the electroweak scale, at temperatures $\sim 100$ GeV, the resulting gravitational wave (GW) signal may fall within the sensitivity range of upcoming space-based detectors such as LISA~\cite{LISA:2017pwj}, Taiji~\cite{Ren:2023yec}, and TianQin~\cite{TianQin:2015yph}. 
Alternatively, if the phase transition takes place during the QCD epoch at temperatures $\sim 200$ MeV, it would generate low-frequency gravitational waves that could leave detectable imprints on pulsar timing arrays (PTAs)~\cite{Yokoyama:2021hsa}. 
Questions regarding the detailed dynamics of the phase transition, such as how it occurs or whether different mechanisms could give rise to distinct types of phase transitions, are beyond the scope of the present work. These issues are part of our planned future research.

As a complementary remark, we can briefly discuss the orbifolding effects of $\mathbb{Z}_{2}$ quotient in the $U(2)$ gauge theory. It is clear that the 4D Minkowski spacetime manifold under consideration supports a spin structure. 
The vanishing of the spin-bordism group, $\Omega_{5}^{\text{spin}}(BU(2))=0$, implies that the exponentiated $\eta$-invariant must be trivial. Thus, this ensures the absence of global anomalies in the 4D $U(2)$ gauge theory~\cite{Davighi:2020bvi}.

%


\section{Summary and conclusions}

As a natural extension of CQFT for the SM particles, we establish the theoretical foundation for the QFT with Wigner degeneracy, which may offer a viable framework for describing DM. In this framework, the Wigner multiplet and its corresponding fields furnish unitary irreducible representations of the extended Poincar\'{e} group, suggesting that DM candidates could naturally arise from such representations.

In this work, we construct a theory of massive Wigner doublet. 
We demonstrate that discrete transformations can map a one-particle state to a superposition of Wigner degenerate states, revealing the intrinsic mixing of Wigner degeneracy.  
To explore this mixing explicitly, we consider a simple yet nontrivial representation where the spatial parity matrix is diagonal, the time-reversal matrix is anti-diagonal, and the charge-conjugation matrix is unitary but otherwise unconstrained.
We propose two reasonable approaches to formulating a QFT for Wigner-degenerate fermions: the doublet construction or a superposition framework. 
In this work, we focus on the doublet construction, where each doublet field consists of two Dirac spinor fields corresponding to distinct Wigner degeneracies. 
This formulation ensures that the Wigner-degenerate fields and their Dirac dual fields manifestly respect both the causality condition and the Lorentz covariance. 
Unlike in CQFT, although a free theory of the Wigner doublet remains $CPT$ invariant, interactions involving the Wigner doublet can explicitly break the $CPT$ invariance.
This $CPT$ violation does not originate from the Lorentz symmetry breaking but rather from the intrinsic structure of the interactions, which mixes the Wigner degeneracies. 
We provide a concrete example of such $CPT$ violation in the context of Yukawa interactions in Section~\ref{sec:W_inter_Yuk}.
However, the $CPT$ invariance can still be restored if it is imposed as a symmetry by hand initially.
We construct a $U(2)$ gauge theory for the Wigner doublet in Section~\ref{sec:W_inter_G}, where the requirement of $CPT$ invariance leads to a specific Wigner symmetry on the gauge fields.
Furthermore, if such a $U(2)$ gauge theory indeed governs the dark sector,
we predict the emergence of a phase transition in the early universe, reflecting the physical distinction introduced by Wigner degeneracy.

There remain several important directions for future research. From a phenomenological perspective, this framework provides new avenues for exploring physics beyond the SM using the doublet formalism. 
The Yukawa and gauge interactions discussed in Section~\ref{sec:W_inter} provide a foundation for further investigations, particularly regarding the implications of $CPT$ violation and phase transition in the early universe. 
Understanding the dynamics of such a phase transition, and whether different mechanisms could produce distinct types of phase transitions, is a compelling topic that lies at the intersection of particle physics and cosmology.
From a theoretical standpoint, an alternative approach involves constructing a QFT based on the superposition of Wigner-degenerate spinor fields. While previous literature \cite{Ahluwalia:2022yvk,Ahluwalia:2023slc} has demonstrated the self-consistency of such a framework using mass-dimension-one fields with Klein-Gordon kinematics, a systematic construction remains an open challenge. Addressing this issue will also be a key objective in our future work.

\acknowledgments
SZ is supported by Natural Science Foundation of China under Grant No.12347101 and No.2024CDJXY022 at Chongqing University.

\appendix

\section*{Appendix}

\section{Notations and conventions}
\label{app:notations_and_conventions}

Throughout the paper, we use the conventions of Ref.~\cite{Peskin:1995ev}.
The 4D Minkowski metric is,
\begin{equation}
\eta_{\mu \nu} = \text{diag}(+1, -1, -1, -1) \, ,
\label{metric_1}
\end{equation}
where $\mu, \nu=0,1, 2, 3$.

The $4\times4$ Dirac matrices are taken in the Weyl representation,
\begin{equation}
\gamma^\mu =
\left[\begin{matrix}
0 & \sigma^\mu \\
\bar{\sigma}^\mu & 0
\end{matrix}\right]
\phantom{000} \text{with} \phantom{000}
\biggl\{
\begin{array}{r c l}
\sigma^\mu &=& \left( \mathds{1}_{2 \times 2}, \sigma^i \right) \, , \\
\bar{\sigma}^\mu &=& \left( \mathds{1}_{2 \times 2}, -\sigma^i \right) \, ,
\end{array}
\label{gamma_1}
\end{equation}
where $\mu=0,1,2,3$ and $\sigma^i$ ($i = 1, 2, 3$) are the three Pauli matrices:
\begin{equation}
\sigma^1 =
\left[\begin{matrix}
0 \ \  & \ \ 1  \\
1 \ \  & \ \ 0  
\end{matrix}\right] \, ,
\phantom{000}
\sigma^2 =
\left[\begin{matrix}
0 \ \  & -i \\
i \ \  & 0
\end{matrix}\right] \, ,
\phantom{000}
\sigma^3 =
\left[\begin{matrix}
1 \ \  &  0 \\
0 \ \  &  -1
\end{matrix}\right] \, ,
\end{equation}
and
\begin{align}
\left\{\sigma^{i}, \sigma^{j} \right\} = 2\delta^{ij} \, , \ \ i,j = 1,2,3 \, .
\end{align}
One has also the chiral operator
\begin{equation}
\gamma^5 = i \gamma^0 \gamma^1 \gamma^2 \gamma^3 =
\left[\begin{matrix}
- \mathds{1}_{2 \times 2} & 0 \\
0 & \mathds{1}_{2 \times 2}
\end{matrix}\right] \, ,
\label{gamma_2}
\end{equation}
and the chiral projection operators
\begin{equation}
P_{L,R} \equiv \dfrac{\mathds{1} \mp \gamma^{5}}{2} \, .
\end{equation}


\section{Standard representations of Poincar\'{e} group and inversions}
\label{sec:common_rep_1}

For a massive particle with mass $m$, we take its standard four-momentum to be $k^{\mu} = \left(m, 0, 0, 0 \right)$~\footnote{This standard four-momentum can't be achieved in the massless case due to the on-shell condition.}. The standard boost taking $k^{\mu}$ to an arbitrary massive momentum $p^{\mu}$ is denoted as $L(p)$ so that $p^{\mu}={L^{\mu}}_{\nu}(p)k^{\nu}$, while $\Lambda$ denotes an arbitrary Lorentz transformation.
The little group is defined as the subgroup of the Lorentz group consisting of the Lorentz transformations $W$ which hold $k^{\mu}$ fixed. Thus, for the massive $k^{\mu}$, the little group is the rotation group $\text{SO}(3)$ in the three dimensional space. 
The little-group transformation $W$ can be decomposed in terms of $\Lambda$ and $L(p)$ as
\begin{equation}
W(\Lambda,p)=L^{-1}(\Lambda p)\Lambda L(p) \, .
\label{W_little_1}
\end{equation}
The unitary operator $U(\Lambda, a)$ in the physical Hilbert space indicates the quantum transformation corresponding to the Poincar\'{e} transformation $(\Lambda, a)$. If there's no translation, $U(\Lambda, 0) \equiv U(\Lambda)$ degrades to a Lorentz transformation. The operators $U(\Lambda, a)$ form a unitary representation of the Poincar\'{e} group in the (infinite dimensional) Hilbert space and satisfy the composition rule
\begin{equation}
U(\Lambda^\prime, a')U(\Lambda, a) = U(\Lambda^\prime \Lambda, \Lambda^\prime a + a') \, ,
\label{U_Lorentz_rep_1}
\end{equation}
and that of the Lorentz group can be derived by setting $a'=a=0$.
We need to clarify that, for physical purposes, what we are actually looking
for is not exactly representations of the Poincar\'{e} (Lorentz) group, but projective representations of
the Poincar\'{e} (Lorentz) group, where an additional phase can exist in the group product.
In the physical quantum Hilbert space, we denote the one-particle states for a massive particle $m>0$ and spin $j$
as $|\p,\sigma\rangle$~\footnote{The index for particle species is hidden.} 
naturally in terms of eigenvectors of the four-momentum and the spin z-projection $\sigma=-j,\cdots, +j$.
They are orthonormalized in a Lorentz invariant convention~\cite{Peskin:1995ev}
\begin{equation}
    \langle \p^{\prime},\sigma^{\prime} |\p,\sigma\rangle 
    = 2 E_{\p} \left(2\pi \right)^{3} \delta^{(3)} \left( \p^{\prime} - \p \right) \delta_{\sigma^{\prime} \sigma} \, .
\label{norm_1p_gen_1}
\end{equation}
The transformation of a one-particle state $|\p,\sigma\rangle$ (with spin $j$) under a homogeneous Lorentz transformation $\Lambda$ will produce an eigenvector of the four-momentum operator with eigenvalue $\Lambda p = (E_{\p_{\Lambda}}, \p_{\Lambda})$. The state after the Lorentz transformation $U(\Lambda)|\p,\sigma\rangle$ can be written as a linear combination of $|\p_{\Lambda},\sigma'\rangle$~(see~\cite[Sec.~2]{Weinberg:1995mt} for details and the Poincar\'{e} algebra)
\begin{equation}
U(\Lambda)|\p,\sigma\rangle
=\sum_{\sigma'}D^{(j)}_{\sigma'\sigma} \left(W(\Lambda,p) \right)|\p_{\Lambda},\sigma'\rangle \, ,
\label{Lorentz_part_1}
\end{equation}
where $D^{(j)}_{\sigma'\sigma} (W)$ is a $2j+1$ (finite) dimensional unitary representation of the little group~\eqref{W_little_1}. 
On the other hand, the one-particle state $|\p,\sigma\rangle$ can be generated by the associated creation operator $a^{\dag}(\p, \sigma)$ acting once on the vacuum state $|0\rangle$~\footnote{The vacuum state is normalized dimensionlessly as $\langle 0|0\rangle = 1$.} 
\begin{equation}
    |\p,\sigma\rangle \equiv \sqrt{2 E_{\p}} \, a^{\dag}(\p, \sigma) |0\rangle \, ,
    \label{1p_gen_1}
\end{equation}
with the on-shell condition $E_{\p} = \sqrt{|\p|^{2}+m^{2}}$ and the canonical commutation (for bosons) or anticommutation (for fermions) relations are realized as
\begin{equation}
\bigl[a(\p, \sigma),a^{\dag}(\p^{\prime},\sigma')\bigr]_{\mp} 
= (2\pi)^{3} \delta^{(3)} \left( \p^{\prime} - \p \right) \delta_{\sigma^{\prime} \sigma} \, ,
\label{can_comm_1}
\end{equation}
with the signs $-$ and $+$ indicating a commutator (for bosons) and an anticommutator (for fermions) respectively.
The vacuum state $|0\rangle$ can be generally destroyed by the annihilation operator, i.e., $a(\boldsymbol{p}, \sigma) |0\rangle = 0$.
One can clearly verify that the canonical quantization relations~\eqref{can_comm_1} are compatible with the normalization of one-particle states in Eq.~\eqref{norm_1p_gen_1}. The Lorentz transformations on the annihilation and creation operators can be derived by those on the states in Eq.~\eqref{Lorentz_part_1}
\begin{align}
U(\Lambda)a^{\dag}(\p, \sigma) U^{-1}(\Lambda)
&=\sqrt{\frac{E_{\p_{\Lambda}}}{E_{\p}}}\sum_{\sigma'}D^{(j)}_{\sigma'\sigma} \left(W(\Lambda,p) \right) a^{\dag}(\p_{\Lambda}, \sigma') 
\nonumber \\
&=\sqrt{\frac{E_{\p_{\Lambda}}}{E_{\p}}}\sum_{\sigma'}D^{(j)*}_{\sigma \sigma'} \left(W^{-1}(\Lambda,p) \right) a^{\dag}(\p_{\Lambda}, \sigma')
\, , 
\label{eq:Lorentz_part_a_1} \\
U(\Lambda)a(\p, \sigma) U^{-1}(\Lambda)
&=\sqrt{\frac{E_{\p_{\Lambda}}}{E_{\p}}}\sum_{\sigma'}D^{(j)*}_{\sigma'\sigma} \left(W(\Lambda,p)\right) a(\p_{\Lambda}, \sigma') 
\nonumber \\
&=\sqrt{\frac{E_{\p_{\Lambda}}}{E_{\p}}}\sum_{\sigma'}D^{(j)}_{\sigma \sigma'} \left(W^{-1}(\Lambda,p) \right)a(\p_{\Lambda}, \sigma')
\, ,
\label{eq:Lorentz_part_a-_1}
\end{align}
where we have used the unitarity of rotation matrices $D^{(j)}_{\sigma'\sigma}$ in the second equality induced by the normalization~\eqref{norm_1p_gen_1} and required that the vacuum state is Lorentz invariant
\begin{equation}
    U(\Lambda)|0\rangle = |0\rangle \, .
\label{eq:Lorentz_vac_1}
\end{equation}
After the proper orthochronous and continuous Lorentz group, we now consider discrete transformations namely, the parity (space inversion) $\mathscr{P}$ and time-reversal $\mathscr{T}$. If they are conserved for the quantum system, there must exist the (projective) (anti)unitary representations corresponding to $\mathscr{P}$ and $\mathscr{T}$, denoted as $P \equiv U(\mathscr{P})$ and $T \equiv U(\mathscr{T})$ respectively, such that they satisfy the multiplication rule with respect to the Poincar\'{e} group
\begin{align}
P U(\Lambda, a) P^{-1} = U(\mathscr{P} \Lambda \mathscr{P}^{-1}, \mathscr{P} a) 
\, , \nonumber \\
T U(\Lambda, a) T^{-1} = U(\mathscr{T} \Lambda \mathscr{T}^{-1}, \mathscr{T} a)
\, .
\label{U_PT_poin_1}
\end{align}
Note that $P$ is linear and unitary while $T$ is antilinear and antiunitary, since we require there is no state of negative energy. We can determine the $P$ and $T$ transformations on the Poincar\'{e} generators using Eq.~\eqref{U_PT_poin_1}:
\begin{align}
P \J P^{-1} &= + \J \, , \label{U_J_P_poin_2} \\
P \K P^{-1} &= - \K \, , \label{U_K_P_poin_2} \\
P \P P^{-1} &= - \P \, , \label{U_P_P_poin_2} \\
P H P^{-1} &= + H \, , \label{U_H_P_poin_2} \\
T \J T^{-1} &= - \J \, , \label{U_J_T_poin_2} \\
T \K T^{-1} &= + \K \, , \label{U_K_T_poin_2} \\
T \P T^{-1} &= - \P \, , \label{U_P_T_poin_2} \\
T H T^{-1} &= + H \, , \label{U_H_T_poin_2}
\end{align}
where $H=P^{0}$ is the Hamiltonian, $\J$ is the three angular momentum (pseudo-) vector, $\mathbf{K}$ is the three boost vector and $\mathbf{P}$ is the three momentum operator. We can then obtain the $P$ and $T$ transformations on the one-particle state $|\p,\sigma\rangle$ 
\begin{align}
P|\p,\sigma\rangle &= \eta_{P}|-\p, \sigma\rangle \, ,
\label{eq:Up}
\\
T|\p,\sigma\rangle &= (-1)^{j-\sigma}\eta_{T}|-\p, -\sigma\rangle \, ,
\label{eq:Ut}
\end{align}
where $\eta_{P,T}$ is the intrinsic phase and independent of the spin-projection $\sigma$, mainly induced by the transformation properties for $\J$ given in Eqs.~\eqref{U_J_P_poin_2}-\eqref{U_J_T_poin_2}. It is remarkable that the time-reversal phase $\eta_{T}$ has no physical influence since it can be eliminated by renormalizing the one-paricle states, which is allowed by the antilinearity of $T$. 
Furthermore, $T^{2}$ has a simple action on the states derived from Eq.~\eqref{eq:Ut}:
\begin{align}
T^{2} |\p,\sigma\rangle 
= (-1)^{2j} |\p,\sigma\rangle 
\, ,
\label{eq:Ut_2}
\end{align}
which has eigenvalues $\pm 1$, only depending on the particle spin $j$. For parity, while we have $\mathscr{P}^{2}=\mathds{1}$, the corresponding transformation in the Hilbert space $P^{2}$ can differ from the identity operator up to a phase $P^{2}|\p,\sigma\rangle=\eta^{2}_{P}|\p,\sigma\rangle$. Since $\eta_{P}$ may be complex, we cannot deduce $\eta_{P} = \pm 1$ directly~\footnote{In general, one can use some internal symmetry operator $I_{P}$ to redefine $PI_{P}$ as a new parity operator such that$\left[PI_{P}\right]^{2}=\boldsymbol{1}$ (see~\cite[Sec.~3.3]{Weinberg:1995mt} for more details).}.

In the same way as the Lorentz transformations in Eqs.~\eqref{eq:Lorentz_part_a_1}-\eqref{eq:Lorentz_part_a-_1}, we can derive the discrete symmetry transformations for the annihilation $a(\p,\sigma)$ and creation operators $a^{\dag}(\p,\sigma)$
\begin{align}
Pa(\p, \sigma) P^{-1} &= \eta_{P}^{*} \, a(-\p, \sigma) \, ,\nonumber\\
P a^{\dag}(\p, \sigma) P^{-1} &= \eta_{P} \, a^{\dag}(-\p, \sigma) \, ,
\label{eq:Up_part_a_1}
\\
T a(\p, \sigma) T^{-1} &= (-1)^{j-\sigma} \eta_{T}^{*} \, a(-\p, -\sigma) \,,\nonumber\\
T a^{\dag}(\p, \sigma) T^{-1} &= (-1)^{j-\sigma}\eta_{T} \, a^{\dag}(-\p, -\sigma) \, .
\label{eq:Ut_part_a_1}
\end{align}

\section{Block-diagonalization of $D(\mathscr{T})$}
\label{sec:T_block_1}

In~\cite[app.~2C]{Weinberg:1995mt}, by exploiting the fact that $T$ is antilinear and antiunitary, Weinberg was able to choose an appropriate basis where $D(\mathscr{T})$ takes the form
\begin{equation}
    D(\mathscr{T})=V\oplus W
\end{equation}
where 
\begin{align}
    V&=\text{diag}(e^{i\theta_{1}},e^{i\theta_{2}},\cdots)\,,\quad\theta_{i}\in\mathbb{R}\\
    W&=W_{1}\oplus W_{2}\oplus\cdots\,,\quad
    W_{i}=\left[\begin{matrix}
        0 & e^{i\phi_{i}/2} \\
        e^{-i\phi_{i}/2} & 0
    \end{matrix}\right],\quad \phi_{i}\in\mathbb{R}.
\end{align}
Since Weinberg's proof is for an arbitrary number of degeneracies, it is difficult to follow. For this reason, it is instructive to review the proof where $D(\mathscr{T})$ is a $2\times2$ unitary matrix. Consider the state $|\p,\sigma,n\rangle'$ whose time-reversal transformation is
\begin{equation}
    T|\p,\sigma,n\rangle'=(-1)^{1/2-\sigma}\sum_{m}{D'}_{mn}(\mathscr{T})|-\p,-\sigma,m\rangle'\,.
\end{equation}
Let $|\p,\sigma,n\rangle'$ be related to $|\p,\sigma,n\rangle$ via a unitary transformation
\begin{equation}
    |\p,\sigma,n\rangle=\sum_{m}\mathscr{U}_{mn}|\p,\sigma,m\rangle'\,.
\end{equation}
Since $T$ is antilinear and antiunitary, the time-reversal matrix $D(\mathscr{T})$ for $|\p,\sigma,n\rangle$ is related to $D'(\mathscr{T})$ via
\begin{equation}
    D(\mathscr{T})=\mathscr{U}^{-1}D'(\mathscr{T})\mathscr{U}^{*}\,.\label{eq:U_trans}
\end{equation}
This transformation is not unitary so $D(\mathscr{T})$ cannot be diagonalized using Eq.~\eqref{eq:U_trans}. However, the basis transformation for $D(\mathscr{T})D^{*}(\mathscr{T})$ is unitary
\begin{equation}
    D(\mathscr{T})D^{*}(\mathscr{T})=\mathscr{U}^{-1}D(\mathscr{T})D^{*}(\mathscr{T})\mathscr{U}\,,
\end{equation}
so it can be diagonalized
\begin{equation}
    D(\mathscr{T})D^{*}(\mathscr{T})=d=\left[\begin{matrix}
        e^{i\frac{\phi}{2}} & 0 \\
        0 & e^{i\phi'/2}
    \end{matrix}\right]\,,\quad\phi,\phi'\in\mathbb{R}\,.\label{eq:d1}
\end{equation}
Take the complex conjugate of Eq.~\eqref{eq:d1}, we find $D^{*}(\mathscr{T})D(\mathscr{T})=d^{-1}$ and
\begin{equation}
    D(\mathscr{T})=\left[\begin{matrix}
        e^{i\phi} & 0 \\
        0 & e^{i\phi'}
    \end{matrix}\right]
    D^{\text{T}}(\mathscr{T})\,.\label{eq:d2}
\end{equation}

We can classify the solutions by setting the two phases equal to or not equal to one. When $e^{i\phi}=e^{i\phi'}=1$, $D(\mathscr{T})$ is symmetric and unitary so it can be written as the exponential of a skew-symmetric matrix which can be diagonalized. Therefore, $D(\mathscr{T})$ can be made to take the form
\begin{equation}
    D(\mathscr{T})=\left[\begin{matrix}
        e^{i\theta_{1}} & 0 \\
        0 & e^{i\theta_{2}}
    \end{matrix}\right]\,,\quad \theta_{i}\in\mathbb{R}\,.
\end{equation}
The case for $e^{i\phi}=1$, $e^{i\phi'}\neq1$ leads to a non invertible $D(\mathscr{T})$ so it is not an admissible solution. Finally, for $e^{\phi}\neq1$, $e^{i\phi'}\neq1$, let us take
\begin{equation}
D(\mathscr{T})=
\left[\begin{matrix}
\alpha \ \  & \ \ \beta  \\
\gamma \ \  & \ \ \delta  
\end{matrix}\right] 
\,,
\end{equation}
and substitute it into Eq.~\eqref{eq:d2}. In this case, the diagonal entries of $D(\mathscr{T})$ vanish and $\gamma=e^{i\phi}\beta$, $e^{i\phi}=e^{-i\phi'}$, so we have
\begin{equation}
D(\mathscr{T})=\left[\begin{matrix}
    0 & \beta \\
    e^{-i\phi}\beta & 0
\end{matrix}\right]\,.
\end{equation}
By performing another basis transformation, we can make the upper right and lower left entries of $D(\mathscr{T})$ to be complex conjugate of each other. This is equivalent to $\beta=e^{i\frac{\phi}{2}}$ so we obtain
\begin{equation}
    D(\mathscr{T})=\left[\begin{matrix}
        0 & e^{i\frac{\phi}{2}} \\
        e^{-i\frac{\phi}{2}} & 0
    \end{matrix}\right]\,,\quad \phi\in\mathbb{R}\,.
\end{equation}

\bibliography{Bibliography}
\bibliographystyle{JHEP}

\end{document}